\documentclass{jfm}
\usepackage{amssymb}
\usepackage{amsmath}
\usepackage{graphicx}
\usepackage{dcolumn}
\usepackage{bm}
\usepackage{epsfig}
\usepackage[T1]{fontenc}
\usepackage{ae,aecompl}
\usepackage{epstopdf, epsfig}
\usepackage{color}
\usepackage{appendix}
\usepackage[export]{adjustbox}
\usepackage{graphicx}
\usepackage{longtable}
\usepackage{epstopdf, epsfig}
\usepackage[latin1]{inputenc}
\usepackage{booktabs}
\usepackage{booktabs}
\usepackage{array}
\usepackage{multirow}

\newcolumntype{C}[1]{>{\centering\arraybackslash}m{#1}}

\shorttitle{Kinetic moment closure and multigradient coupling in high-order TNE}
\shortauthor{Lai, Guo, Gan, Yang, Liu and Lin}
\affiliation{
\aff{1}School of Mathematics and Statistics \& Key Laboratory of Analytical Mathematics and Applications (Ministry of Education) \& Fujian Key Laboratory of Analytical Mathematics and Applications (FJKLAMA) \& Center for Applied Mathematics of Fujian Province (FJNU), Fujian Normal University, Fuzhou 350117, P. R. China
\aff{2}Hebei Key Laboratory of Trans-Media Aerial Underwater Vehicle, North China Institute of Aerospace Engineering, Langfang 065000, P. R. China
\aff{3}School of Energy and Safety Engineering, Tianjin Chengjian University, Tianjin 300384, China
\aff{4}Laoshan Laboratory, Qingdao 266237, China
\aff{5}State Key Laboratory of Earth System Numerical Modeling and Application, Institute of Atmospheric Physics, Chinese Academy of Sciences, Beijing 100029, China
\aff{6}College of Earth and Planetary Sciences, University of Chinese Academy of Sciences, Beijing 100049, China
}
\begin{document}

\title{High-order thermodynamic nonequilibrium in three-dimensional
compressible flows: Kinetic moment closure and multigradient coupling}
\author{Huilin Lai\aff{1}, Qinghong Guo\aff{1}, Yanbiao Gan\aff{2}%
\corresp{\email{Gan@nciae.edu.cn}}, Bin Yang\aff{3}, Hailong Liu\aff{4}, Pengfei Lin\aff{5,6} }
\maketitle

\begin{abstract}
High-order thermodynamic nonequilibrium (TNE) in three-dimensional
compressible flows reflects the breakdown of low-order kinetic moment
closure in strong-gradient regions. Using Chapman--Enskog analysis, we
identify the kinetic moment constraints required to describe third-order
TNE. The analysis yields the third-order constitutive relations and evolution equations for the viscous stress $\bm{\Delta}_{2}^{*}$ and heat flux $\bm{\Delta}_{3,1}^{*}$, together with second-order expressions for their associated higher-order fluxes, $\bm{\Delta}_{3}^{*}$ and $\bm{\Delta}_{4,2}^{*}$.
These constraints enable the construction of a
three-dimensional super-Burnett-level discrete Boltzmann model with 91
discrete velocities.
The resulting D3V91 model reproduces shock-tube wave structures and resolves high-order TNE contributions that lower-order DBMs do not capture reliably.
These results demonstrate that high-order TNE has a multigradient, rather
than single-gradient, origin. For the four TNE quantities considered here,
odd-order central moments, including the heat flux
$\bm{\Delta}_{3,1}^{*}$ and the viscous-stress flux
$\bm{\Delta}_{3}^{*}$, are primarily governed by temperature gradients,
whereas even-order central moments, including the viscous stress
$\bm{\Delta}_{2}^{*}$ and the heat-flux-related flux
$\bm{\Delta}_{4,2}^{*}$, are dominated by velocity gradients.
These leading-gradient dependences are not exclusive; they are substantially modified by density gradients, secondary gradients and transition-layer widths through higher-order derivative terms, gradient products and cross-couplings.
When the secondary contributions become comparable to the
leading-gradient terms, the nonequilibrium response transitions from a near-linear
regime to an approximately exponential regime.
This work establishes a super-Burnett-level DBM framework that treats kinetic moment closure and multigradient coupling consistently, providing a basis for resolving and interpreting high-order TNE in three-dimensional compressible flows.

\end{abstract}




\section{\label{sec:level1} Introduction}

High-speed compressible flows commonly contain strong-gradient structures,
including shock waves, rarefaction waves, contact discontinuities and shear
layers. Across these structures, density, temperature and velocity may vary
over length scales comparable to the kinetic relaxation scale, so local
kinetic effects become appreciable. The molecular velocity distribution can
then depart markedly from local equilibrium, producing thermodynamic
nonequilibrium (TNE) effects that cannot be inferred from conserved
macroscopic variables alone \citep{cercignani2000rarefied,42Xu2022}. Such
effects are central to hypersonic, reacting, detonating and shock-driven
flows, as well as to spacecraft re-entry aerothermodynamics and
high-energy-density physics
\citep{5Courant1999SSBM,wang2024experimental,nagnibeda2009non,7Hosseini2024PECS,zhou2017rayleigh,liang2023review,ivanov1998computational,chen2026observation}.
Resolving these effects is essential because nonequilibrium transport
controls shock thickness, temperature overshoot, heat flux, viscous
dissipation and entropy production, and mediates the feedback between
kinetic relaxation and macroscopic flow evolution.

The kinetic information relevant to TNE is not contained in the conserved
fields alone. Density, velocity and temperature specify the macroscopic
state but do not determine the nonequilibrium part of the distribution
function. Non-conserved kinetic moments, including viscous stress, heat flux
and their higher-order fluxes, quantify local transport, dissipation and
relaxation. They therefore link mesoscale structures directly to
nonequilibrium transport mechanisms. As nonequilibrium strengthens,
conserved variables and low-order fluxes no longer provide a sufficient
description of the system state. Higher-order non-conserved moments then
become sensitive probes of small-scale structures, fast kinetic modes and
strongly coupled components of the distribution function. Resolving
high-order TNE is therefore essential both for kinetic modelling and for
identifying the driving mechanisms and coupling pathways of nonequilibrium
transport \citep{50Gan2022JFM,xu2024advances}.

The Navier--Stokes--Fourier (NSF) equations describe gas flows effectively
near local equilibrium. Their validity relies on scale separation: the
molecular mean free path must be much smaller than the macroscopic length
scale, and the kinetic relaxation time must be much shorter than the
characteristic flow time. When this separation weakens, non-local kinetic
effects become important and the NSF description loses validity
\citep{bird1994molecular,16Struchtrup2005}. At the NSF level, viscous stress
and heat flux are linearly related to local velocity and temperature
gradients. These first-order constitutive relations are adequate only for
weak nonequilibrium with smoothly varying gradients. In strong-gradient
regions, stress and heat flux may also depend on density variation, velocity
compression, temperature curvature, relaxation inhomogeneity and products of
different gradients. These contributions involve higher-order kinetic
moments and therefore cannot be represented by a first-order closure
\citep{chapman1990mathematical}. A low-order model may therefore reproduce
the main wave pattern while still failing to describe local nonequilibrium
transport inside shock layers, near contact discontinuities and around
extrema of macroscopic gradients \citep{mott1951solution,48Gan2018PRE}.

Burnett-level descriptions extend the NSF approximation by retaining
second-order terms in the Chapman--Enskog expansion
\citep{garcia2008beyond}. This extension accounts for part of the non-local
transport, but remains insufficient when third-order TNE contributions become
non-negligible. In the Chapman--Enskog framework, the Euler, NSF, Burnett and
super-Burnett levels correspond to zeroth-, first-, second- and third-order
approximations in the Knudsen number, respectively. Here, super-Burnett level
is used in this kinetic sense: it refers to the moment constraints required
for third-order TNE, not to the direct solution of the classical
super-Burnett transport equations. A super-Burnett-level description must
therefore connect viscous stress and heat flux to their associated
higher-order fluxes and represent their evolution consistently. This
requirement leads directly to the kinetic moment closure problem.

This closure problem is central to high-order discrete kinetic modelling. A
discrete kinetic model does not resolve a nonequilibrium quantity simply by
recovering the conserved moments or reproducing macroscopic profiles. To
describe a target TNE quantity, the discrete velocity set and the
corresponding equilibrium distribution must satisfy the kinetic moment
relations associated with that quantity and with the prescribed
approximation order. Otherwise, the model may predict density, velocity and
temperature accurately while giving incorrect non-conserved moments, such as
high-order heat fluxes, viscous-stress fluxes and their extrema.
Thus, constructing a high-order discrete Boltzmann model is not merely a
matter of increasing the number of discrete velocities; it requires
identifying the moment constraints associated with the target
nonequilibrium quantities and designing a discrete velocity set that
satisfies them.

Strong nonequilibrium also raises a mechanistic question: how do multiple
macroscopic gradients jointly drive high-order TNE quantities? These
quantities are generally not governed by a single gradient. Different
gradients may dominate different kinetic moments: temperature gradients
primarily drive energy-transport moments, whereas velocity gradients
dominate momentum-transport moments. Density gradients, secondary gradients
and transition-layer widths further modify these leading dependences through
gradient products, curvature terms and cross-couplings. Such coupling is
particularly important in shock and interface problems, where density,
velocity and temperature vary simultaneously over comparable spatial scales.
High-order TNE should therefore be viewed as a multigradient-coupled kinetic
response, rather than as a one-to-one correspondence between a
nonequilibrium quantity and a macroscopic gradient.

Kinetic approaches provide a natural framework for analysing these effects.
Direct Boltzmann solvers, simplified collision models and multiscale unified
frameworks have been widely used to study nonequilibrium gas flows
\citep{17Yin2026SJSC,18Wu2017CAS,hosseini2023lattice,21Guo2026AA,23Lu2026JOCP,31Fei2023JOFM}.
Among these approaches, the discrete Boltzmann method (DBM) is particularly
useful for TNE analysis because it provides non-conserved kinetic moments as
well as macroscopic fields
\citep{50Gan2022JFM,xu2024advances,37Zhang2026FOP,lin2016double,lin2019discrete}.
These moments allow viscous stress, heat flux, higher-order fluxes and other
TNE measures to be extracted and compared with analytical constitutive
relations. DBM therefore serves both as a flow solver and as a kinetic
diagnostic framework for nonequilibrium transport.

Existing three-dimensional DBM formulations, however, remain mainly at the
NSF \citep{43Gan2018SPS,44Ji2021AA,45Ji2022JOCP} or Burnett level
\citep{47Guo2025POF}. Although these models can represent some lower-order
nonequilibrium effects, they do not provide the kinetic moment closure
required for third-order TNE. In particular, a Burnett-level model may
describe part of the second-order nonequilibrium response, but it cannot
reliably resolve the third-order constitutive structure of viscous stress
and heat flux or the second-order behaviour of their associated higher-order
fluxes. This restriction limits the analysis of how different gradients
drive, amplify or suppress high-order TNE quantities, and motivates the
development of a three-dimensional super-Burnett-level kinetic closure
within the DBM framework.

Here we construct such a closure and use it to analyse high-order TNE in three-dimensional compressible flows. Starting from the Chapman--Enskog expansion, we identify the
kinetic moment constraints required for third-order TNE. We derive third-order
constitutive relations and evolution equations for viscous stress and heat flux, together
with second-order expressions for their associated higher-order fluxes. We then construct a 91-velocity discrete velocity set, denoted D3V91, that satisfies these constraints. We first test the model for macroscopic consistency using shock-tube problems and then use it to analyse second- and third-order non-conserved kinetic moments driven by temperature, velocity and density gradients and by the transition-layer width.
Unlike lower-order models such as D3V55 \citep{47Guo2025POF}, the proposed D3V91 model is designed to enforce the super-Burnett-level moment constraints, rather than merely to increase the number of discrete velocities. This provides a
three-dimensional kinetic closure that links super-Burnett-level moment constraints to
the multigradient coupling of high-order TNE.

The remainder of this paper is organised as follows. Section~2 derives the kinetic moment relations required for third-order TNE and presents the analytical expressions and evolution equations for the relevant nonequilibrium quantities. Section~3 constructs the D3V91 discrete velocity set and the corresponding three-dimensional super-Burnett-level DBM, and validates the model using one-dimensional benchmark problems. Section~4 analyses the second- and third-order nonequilibrium quantities, including $\bm{\Delta}_3^*$, $\bm{\Delta}_{4,2}^*$, $\bm{\Delta}_2^*$ and $\bm{\Delta}_{3,1}^*$, with emphasis on multigradient coupling mechanisms. Section~5 concludes the paper and outlines future directions.

\section{A three-dimensional super-Burnett-level discrete Boltzmann model}
\label{Model}

This section develops a three-dimensional discrete Boltzmann model for
resolving third-order TNE. The construction has two objectives. The first is
to identify, through the CE expansion, the kinetic moment
constraints required to describe the third-order constitutive relations and
evolution equations for the viscous stress $\bm{\Delta}_{2}^{*}$ and heat
flux $\bm{\Delta}_{3,1}^{*}$, together with the second-order expressions for
their associated higher-order fluxes, namely the viscous-stress flux
$\bm{\Delta}_{3}^{*}$ and the heat-flux-related flux
$\bm{\Delta}_{4,2}^{*}$. The second is to derive these TNE quantities. In this sense,
the term super-Burnett-level denotes the kinetic moment closure required for
third-order TNE, rather than a direct numerical solution of the classical
super-Burnett transport equations.

\subsection{Kinetic moment constraints for third-order TNE}

The discrete Boltzmann equation is written as
\begin{equation}  \label{e1}
\frac{\partial f_i}{\partial t} + \mathbf{v}_i \cdot \frac{\partial f_i}{%
\partial \mathbf{r}} =-\frac{1}{\tau}[f_i-f_i^{(0)}],
\end{equation}
where $i=1,\ldots,N$ indexes the discrete velocities, $f_i$ is the discrete
distribution function, $f_i^{(0)}$ is the discrete equilibrium distribution,
$\mathbf{v}_i$ is the discrete velocity, and $\tau$ is the relaxation time.

Velocity discretisation is introduced not to preserve the continuous
distribution function pointwise, but to retain the kinetic moments
required by the target hydrodynamic and TNE descriptions.
These moments
carry the physical information needed to describe macroscopic fields and
nonequilibrium responses. The discrete velocity set and the corresponding
equilibrium distribution must therefore reproduce the required continuous
moment relations,
\begin{equation}  \label{e2}
\bm{\Phi}^{\prime }= \int f \bm{\Psi}^{\prime }(\mathbf{v}, \eta) d \mathbf{v%
} = \sum_i f_i \bm{\Psi}^{\prime }(\mathbf{v}_i, \eta_i),
\end{equation}
where $\bm{\Psi}^{\prime }$ denotes the set of moment basis functions. The
order of this basis determines the highest-order nonequilibrium information
that the model can represent.

To determine these constraints, we apply the CE expansion to
Eq.~(\ref{e1}). The distribution function, time derivative and spatial
derivative are expanded as
\begin{equation}  \label{e3}
f_i = f_i^{(0)} + f_i^{(1)} + f_i^{(2)} + f_i^{(3)} + f_i^{(4)} + \cdots,
\end{equation}
\begin{equation}  \label{e4}
\partial_t = \partial_{t_1} + \partial_{t_2} + \partial_{t_3} +
\partial_{t_4} + \cdots,
\end{equation}
\begin{equation}  \label{e5}
\bm{\nabla} = \bm{\nabla}_1 .
\end{equation}
Here $f_i^{(j)}$ denotes the $j$th-order departure from equilibrium, and $\partial_{t_j}$ denotes differentiation on the corresponding time scale.

Substituting Eqs.~(\ref{e3})--(\ref{e5}) into Eq.~(\ref{e1}) gives
\begin{equation}  \label{e6}
\begin{aligned} & \bigl( \partial_{t_1} + \partial_{t_2} + \partial_{t_3} +
\partial_{t_4} + \cdots \bigr) \bigl( f_i^{(0)} + f_i^{(1)} + f_i^{(2)} +
f_i^{(3)} + f_i^{(4)} + \cdots \bigr) \\ &\quad + \mathbf{v}_i \cdot
\bm{\nabla}_1 \bigl( f_i^{(0)} + f_i^{(1)} + f_i^{(2)} + f_i^{(3)} +
f_i^{(4)} + \cdots \bigr) \\ & = -\frac{1}{\tau} \bigl(
f_i^{(1)}+f_i^{(2)}+f_i^{(3)}+f_i^{(4)}+\cdots \bigr). \end{aligned}
\end{equation}
Collecting terms at the same asymptotic order yields
\begin{equation}  \label{e7}
f_i^{(1)} = -\tau \left[\partial_{t_1} f_i^{(0)} + \bm{\nabla}_1 \cdot
(f_i^{(0)} \mathbf{v}_i)\right],
\end{equation}
\begin{equation}  \label{e8}
\begin{aligned} f_i^{(2)} = - & \tau \left[\partial_{t_2} f_i^{(0)} +
\partial_{t_1}f_i^{(1)} + \bm{\nabla}_1 \cdot (f_i^{(1)}
\mathbf{v}_i)\right] \\ = - &\tau \partial_{t_2} f_i^{(0)} + \tau^2
\partial_{t_1}^2 f_i^{(0)} + \tau^2 \partial_{t_1} \left[\bm{\nabla}_1 \cdot
(f_i^{(0)} \mathbf{v}_i)\right] \\ + & \tau^2 \bm{\nabla}_1 \cdot
\left[\partial_{t_1} f_i^{(0)} \mathbf{v}_i + \bm{\nabla}_1 \cdot (f_i^{(0)}
\mathbf{v}_i \mathbf{v}_i)\right], \end{aligned}
\end{equation}
and
\begin{equation}  \label{e9}
\begin{aligned} f_i^{(3)} & =-\tau\left[\partial_{t_3}
f_i^{(0)}+\partial_{t_2} f_i^{(1)} +\partial_{t_1} f_i^{(2)}+\bm{\nabla}_1
\cdot\left(f_i^{(2)} \mathbf{v}_i\right)\right] \\ & =-\tau \partial_{t_3}
f_i^{(0)}+\tau^2 \partial_{t_2}\left[\partial_{t_1} f_i^{(0)} +\bm{\nabla}_1
\cdot\left(f_i^{(0)} \mathbf{v}_i\right)\right] +\tau^2
\partial_{t_1}\left(\partial_{t_2} f_i^{(0)}\right)-\tau^3 \partial_{t_1}^3
f_i^{(0)}\\ &\quad -\tau^3 \partial_{t_1}^2\left[\bm{\nabla}_1
\cdot\left(f_i^{(0)} \mathbf{v}_i\right)\right] -\tau^3
\partial_{t_1}\left\{\bm{\nabla}_1 \cdot \left[\partial_{t_1} f_i^{(0)}
\mathbf{v}_i +\bm{\nabla}_1 \cdot\left(f_i^{(0)} \mathbf{v}_i
\mathbf{v}_i\right)\right]\right\}\\ &\quad +\tau^2 \bm{\nabla}_1
\cdot\left[\partial_{t_2}\left(f_i^{(0)} \mathbf{v}_i\right)\right] -\tau^3
\bm{\nabla}_1 \cdot\left[\partial_{t_1}^2\left(f_i^{(0)}
\mathbf{v}_i\right)\right]\\ &\quad -\tau^3 \bm{\nabla}_1 \cdot
\partial_{t_1}\left[\bm{\nabla}_1 \cdot\left(f_i^{(0)} \mathbf{v}_i
\mathbf{v}_i\right)\right] -\tau^3 \bm{\nabla}_1^2 \cdot\left[\partial_{t_1}
f_i^{(0)} \mathbf{v}_i \mathbf{v}_i +\bm{\nabla}_1 \cdot\left(f_i^{(0)}
\mathbf{v}_i \mathbf{v}_i \mathbf{v}_i\right)\right]. \end{aligned}
\end{equation}

Equations~(\ref{e7})--(\ref{e9}) show that $f_i^{(3)}$ can be expressed
in terms of $f_i^{(0)}$ and its derivatives, but involves higher-order
velocity polynomials than those required at the NSF or Burnett level. A DBM designed to resolve third-order TNE must therefore impose sufficiently high-order equilibrium moment constraints.
Accordingly, we choose the moment basis as
$\bm{\Psi}(\mathbf{v}, \eta)=
[ 1, \mathbf{v}, \frac{1}{2}(v^2+\eta^2), \mathbf{vv}, \frac{1}{2}%
(v^2+\eta^2)\mathbf{v}, \mathbf{vvv}, \frac{1}{2}(v^2+\eta^2)\mathbf{vv},
\mathbf{vvvv}, \frac{1}{2}(v^2+\eta^2)\mathbf{vvv}, \mathbf{vvvvv}, \frac{1}{%
2}(v^2+\eta^2)\mathbf{vvvv} ]$. This basis provides the equilibrium moment
relations needed for third-order TNE closure. The corresponding equilibrium
moments are
\begin{equation}  \label{e10}
\mathbf{M}_0=\sum_i f_i^{(0)}=\rho,
\end{equation}
\begin{equation}  \label{e11}
\mathbf{M}_1=\sum_i f_i^{(0)} \mathbf{v}_i=\rho \mathbf{u},
\end{equation}
\begin{equation}  \label{e12}
\mathbf{M}_{2,0}=\sum_i \frac{1}{2} f_i^{(0)}(v_i^2+\eta_i^2) =\frac{1}{2}
\rho\left[(n+3) R T+u^2\right],
\end{equation}
\begin{equation}  \label{e13}
\mathbf{M}_2=\sum_i f_i^{(0)} \mathbf{v}_i \mathbf{v}_i =\rho\left(R T
\mathbf{I}+\mathbf{u u}\right),
\end{equation}
\begin{equation}  \label{e14}
\mathbf{M}_{3,1}=\sum_i \frac{1}{2} f_i^{(0)}(v_i^2+\eta_i^2) \mathbf{v}_i =%
\frac{1}{2} \rho \mathbf{u}\left[(n+5) R T+u^2\right],
\end{equation}
\begin{equation}  \label{e15}
\begin{aligned} \mathbf{M}_3 =\sum_i f_i^{(0)} \mathbf{v}_i \mathbf{v}_i
\mathbf{v}_i =\rho\left[ R T\left(u_\alpha \delta_{\beta \gamma} + u_\beta
\delta_{\alpha \gamma} + u_\gamma \delta_{\alpha \beta}\right)
\mathbf{e}_\alpha \mathbf{e}_\beta \mathbf{e}_\gamma +\mathbf{u u u}
\right], \end{aligned}
\end{equation}
\begin{equation}  \label{e16}
\begin{aligned} \mathbf{M}_{4,2} &=\sum_i \frac{1}{2}
f_i^{(0)}\left(v_i^2+\eta_i^2\right) \mathbf{v}_i \mathbf{v}_i \\
&=\frac{1}{2} \rho\left[(n+5) R^2 T^2+R T u^2\right] \mathbf{I} +\frac{1}{2}
\rho\left[(n+7) R T+u^2\right] \mathbf{u u}, \end{aligned}
\end{equation}
\begin{equation}\label{e17}
\begin{aligned}
\mathbf{M}_4&=\sum_i f_i^{(0)} \mathbf{v}_i \mathbf{v}_i \mathbf{v}_i \mathbf{v}_i=\rho\left(R^2 T^2\left(\delta_{\alpha \beta} \delta_{\gamma \lambda}+\delta_{\alpha \gamma} \delta_{\beta \lambda}+\delta_{\alpha \lambda} \delta_{\beta \gamma}\right) \mathbf{e}_\alpha \mathbf{e}_\beta \mathbf{e}_\gamma \mathbf{e}_\lambda+ R T\left(u_\alpha u_\beta \delta_{\gamma \lambda}\right.\right.\\
&\left.\left.+u_\alpha u_\gamma \delta_{\beta \lambda} +u_\alpha u_\lambda \delta_{\beta \gamma}+u_\beta u_\gamma \delta_{\alpha \lambda}+u_\beta u_\lambda \delta_{\alpha \gamma}+u_\gamma u_\lambda \delta_{\alpha \beta}\right) \mathbf{e}_\alpha \mathbf{e}_\beta \mathbf{e}_\gamma \mathbf{e}_\lambda+\mathbf{u u u u}\right),
\end{aligned}
\end{equation}
\begin{equation}\label{e18}
\begin{aligned}
\mathbf{M}_{5,3}&=\sum_i \frac{1}{2} f_i^{(0)}\left(v_i^2+\eta_i^2\right) \mathbf{v}_i \mathbf{v}_i \mathbf{v}_i=\rho\left[\left(\frac{n+9}{2} R T+\frac{u^2}{2}\right) \mathbf{u u u}\right.\\
&\left.+\left(\frac{n+7}{2} R T+\frac{u^2}{2}\right) R T\left(u_\alpha \delta_{\beta \gamma} +u_\beta \delta_{\alpha \gamma}+u_\gamma \delta_{\alpha \beta}\right) \mathbf{e}_\alpha \mathbf{e}_\beta \mathbf{e}_\gamma\right],
\end{aligned}
\end{equation}
\begin{equation}\label{e19}
\begin{aligned}
\mathbf{M}_5&=\sum_i f_i^{(0)} \mathbf{v}_i \mathbf{v}_i \mathbf{v}_i \mathbf{v}_i \mathbf{v}_i =\rho\left[R ^ { 2 } T ^ { 2 } \left(u_\alpha \delta_{\beta \gamma} \delta_{\lambda \varepsilon}+u_\alpha \delta_{\beta \lambda} \delta_{\gamma \varepsilon}+u_\alpha \delta_{\beta \varepsilon} \delta_{\gamma \lambda}+u_\beta \delta_{\alpha \gamma} \delta_{\lambda \varepsilon}\right.\right. \\
& +u_\beta \delta_{\alpha \lambda} \delta_{\gamma \varepsilon}+u_\beta \delta_{\alpha \varepsilon} \delta_{\gamma \lambda}+u_\gamma \delta_{\alpha \beta} \delta_{\lambda \varepsilon}+u_\gamma \delta_{\alpha \lambda} \delta_{\beta \varepsilon}+u_\gamma \delta_{\alpha \varepsilon} \delta_{\beta \lambda}+u_\lambda \delta_{\alpha \beta} \delta_{\gamma \varepsilon}+u_\lambda \delta_{\alpha \gamma} \delta_{\beta \varepsilon}\\
&\left.+u_\lambda \delta_{\alpha \varepsilon} \delta_{\beta \gamma}+u_{\varepsilon} \delta_{\alpha \beta} \delta_{\gamma \lambda}+u_{\varepsilon} \delta_{\alpha \gamma} \delta_{\beta \lambda}+u_{\varepsilon} \delta_{\alpha \lambda} \delta_{\beta \gamma}\right) \mathbf{e}_\alpha \mathbf{e}_\beta \mathbf{e}_\gamma \mathbf{e}_\lambda \mathbf{e}_{\varepsilon} +R T\left(u_\gamma u_\lambda u_{\varepsilon} \delta_{\alpha \beta}\right.\\
& \left.\left.+u_\beta u_\lambda u_{\varepsilon} \delta_{\alpha \gamma}+u_\beta u_\gamma u_{\varepsilon} \delta_{\alpha \lambda}+u_\beta u_\lambda u_\gamma \delta_{\alpha \varepsilon}+u_\alpha u_\lambda u_{\varepsilon} \delta_{\beta \gamma}+u_\alpha u_\gamma u_{\varepsilon} \delta_{\beta \lambda}+u_\alpha u_\lambda u_\gamma \delta_{\beta \varepsilon} \right.\right.\\
&\left.\left.+u_\alpha u_\beta u_{\varepsilon} \delta_{\gamma \lambda}+u_\alpha u_\beta u_\lambda \delta_{\gamma \varepsilon}+u_\alpha u_\beta u_\gamma \delta_{\lambda \varepsilon}\right) \mathbf{e}_\alpha \mathbf{e}_\beta \mathbf{e}_\gamma \mathbf{e}_\lambda \mathbf{e}_{\varepsilon}+\mathbf{u u u u u}\right],
\end{aligned}
\end{equation}
\begin{equation}\label{e20}
\begin{aligned}
& \mathbf{M}_{6,4}=\sum_i \frac{1}{2} f_i^{(0)}\left(v_i^2+\eta_i^2\right) \mathbf{v}_i \mathbf{v}_i \mathbf{v}_i \mathbf{v}_i \\
& =\rho\left[\left(\frac{n+11}{2} R T+\frac{u^2}{2}\right) \mathbf{u u u u}+\left(\frac{n+9}{2} R T+\frac{u^2}{2}\right) R T\left(u_\alpha u_\beta \delta_{\gamma \lambda}\right.\right. \\
& \left.+u_\alpha u_\gamma \delta_{\beta \lambda}+u_\alpha u_\lambda \delta_{\beta \gamma}+u_\beta u_\gamma \delta_{\alpha \lambda}+u_\beta u_\lambda \delta_{\alpha \gamma}+u_\gamma u_\lambda \delta_{\alpha \beta}\right) \mathbf{e}_\alpha \mathbf{e}_\beta \mathbf{e}_\gamma \mathbf{e}_\lambda \\
& \left.+\left(\frac{n+7}{2} R T+\frac{u^2}{2}\right) R^2 T^2\left(\delta_{\alpha \beta} \delta_{\gamma \lambda}+\delta_{\alpha \gamma} \delta_{\beta \lambda}+\delta_{\alpha \lambda} \delta_{\beta \gamma}\right) \mathbf{e}_\alpha \mathbf{e}_\beta \mathbf{e}_\gamma \mathbf{e}_\lambda\right].
\end{aligned}
\end{equation}
Together, Eqs.~(\ref{e10})--(\ref{e20}) provide 91 independent equilibrium moment constraints. These constraints define the moment space of the D3V91 model.

These constraints can be written in matrix form as
\begin{equation}  \label{eq:matrix-form}
\bm{\Phi} = \mathbf{C} \mathbf{f}^{(0)},
\end{equation}
where
\begin{equation*}
\bm{\Phi} = (\mathbf{M}_0, \mathbf{M}_1, \mathbf{M}_{2,0}, \mathbf{M}_{2},
\mathbf{M}_{3,1}, \mathbf{M}_{3}, \mathbf{M}_{4,2}, \mathbf{M}_{4}, \mathbf{M%
}_{5,3}, \mathbf{M}_{5}, \mathbf{M}_{6,4})^T
\end{equation*}
is the vector of independent equilibrium moments. The matrix $\mathbf{C}$ maps the discrete equilibrium distribution to this moment vector. For the D3V91
model, $\mathbf{C}$ is a $91\times91$ square matrix. Provided that $\mathbf{C}$ is invertible, the equilibrium distribution is obtained from
\begin{equation}  \label{eq:feq-matrix}
\mathbf{f}^{(0)} = \mathbf{C}^{-1} \bm{\Phi}.
\end{equation}
Thus, the D3V91 velocity set is defined by the super-Burnett-level moment constraints required for third-order TNE. Its role is not merely to increase the number of discrete velocities, but to ensure that the target non-conserved kinetic moments are represented consistently.

\subsection{Derivation of high-order TNE effects}

Taking the zeroth-, first- and second-order moments of Eq.~(\ref{e1}) gives
the generalized hydrodynamic equations
\begin{equation}  \label{eq:mass}
\partial_t \rho + \bm{\nabla} \cdot(\rho \mathbf{u})=0,
\end{equation}
\begin{equation}  \label{eq:momentum}
\partial_t (\rho \mathbf{u}) + \bm{\nabla} \cdot(\rho \mathbf{u u}+P \mathbf{%
I}+\bm{\Delta}_2^*)=0,
\end{equation}
\begin{equation}  \label{eq:energy}
\partial_t(\rho e) + \bm{\nabla} \cdot[(\rho e+P) \mathbf{u} +\bm{\Delta}%
_2^* \cdot \mathbf{u} +\bm{\Delta}_{3,1}^*]=0,
\end{equation}
where $P=\rho RT$, $e=c_v T+u^2/2$, and $c_v=(n+3)R/2$. Here
$\bm{\Delta}_2^*$ denotes the non-organized momentum flux, or viscous
stress, and $\bm{\Delta}_{3,1}^*$ denotes the non-organized energy flux, or
heat flux.

We next define the generalized nonequilibrium moments used to quantify TNE.
The non-central moment is
\begin{equation}  \label{e28}
\begin{aligned} \bm{\Delta}_{m,n} = \mathbf{M}_{m,n}(f_i-f_i^{(0)}) = \sum_i
\left(\frac{1}{2}\right)^{1-\delta_{mn}} (f_i-f_i^{(0)})
\underbrace{\mathbf{v}_i \mathbf{v}_i \cdots \mathbf{v}_i}_{n}
(v_i^2+\eta_i^2)^{(m-n)/2}, \end{aligned}
\end{equation}
and the corresponding central moment is
\begin{equation}  \label{e29}
\begin{aligned} \bm{\Delta}_{m,n}^* = \mathbf{M}_{m,n}^*(f_i-f_i^{(0)}) =
\sum_i \left(\frac{1}{2}\right)^{1-\delta_{mn}} (f_i-f_i^{(0)})
\underbrace{\mathbf{v}_i^* \mathbf{v}_i^* \cdots \mathbf{v}_i^*}_{n}
(v_i^{*2}+\eta_i^2)^{(m-n)/2}. \end{aligned}
\end{equation}
The contribution associated with the $j$th-order departure $f_i^{(j)}$ is
defined as
\begin{equation}  \label{e30}
\begin{aligned} \bm{\Delta}_{m,n}^{*(j)} = \mathbf{M}_{m,n}^*(f_i^{(j)}) =
\sum_i \left(\frac{1}{2}\right)^{1-\delta_{mn}} f_i^{(j)}
\underbrace{\mathbf{v}_i^* \mathbf{v}_i^* \cdots \mathbf{v}_i^*}_{n}
(v_i^{*2}+\eta_i^2)^{(m-n)/2}, \end{aligned}
\end{equation}
where $\delta_{mn}$ is the Kronecker delta and
$\mathbf{v}_i^*=\mathbf{v}_i-\mathbf{u}$. When $m=n$,
$\bm{\Delta}_{m,n}^*$ is abbreviated as $\bm{\Delta}_m^*$. Replacing
$\mathbf{v}_i^*$ with $\mathbf{v}_i$ gives the corresponding non-central
moment.

The evolution equations for the non-conserved central moments are obtained
by taking higher-order moments of the discrete Boltzmann equation. For a
central moment of order $m$, the evolution equation is
\begin{equation}\label{e26}
\begin{aligned}
\partial_t \bm{\Delta}_m^*
+\partial_t \mathbf{M}_m^*(f_i^{(0)})
+\bm{\nabla}\cdot
\big[
&\mathbf{M}_{m+1}^*(f_i^{(0)})
+\mathbf{M}_m^*(f_i^{(0)})\mathbf{u}  \\
&+\bm{\Delta}_{m+1}^*
+\bm{\Delta}_m^*\mathbf{u}
\big]
=
-\frac{1}{\tau}\bm{\Delta}_m^* .
\end{aligned}
\end{equation}
The corresponding contracted central moment satisfies
\begin{equation}\label{e27}
\begin{aligned}
\partial_t \bm{\Delta}_{m,n}^*
+\partial_t \mathbf{M}_{m,n}^*(f_i^{(0)})
+\bm{\nabla}\cdot
\big[
&\mathbf{M}_{m+1,n+1}^*(f_i^{(0)})
+\mathbf{M}_{m,n}^*(f_i^{(0)})\mathbf{u} \\
&+\bm{\Delta}_{m+1,n+1}^*
+\bm{\Delta}_{m,n}^*\mathbf{u}
\big]
=
-\frac{1}{\tau}\bm{\Delta}_{m,n}^* .
\end{aligned}
\end{equation}
Equations~(\ref{e26}) and~(\ref{e27}) show that each non-conserved TNE
quantity evolves through its associated higher-order flux. Thus, viscous
stress and heat flux cannot be closed independently once high-order TNE
effects are retained. Their evolution requires the corresponding higher-order
fluxes, $\bm{\Delta}_{3}^{*}$ and $\bm{\Delta}_{4,2}^{*}$, which provides
the kinetic origin of the moment-closure requirement.

We focus on four TNE quantities. The viscous stress
$\bm{\Delta}_{2}^{*}$ and heat flux $\bm{\Delta}_{3,1}^{*}$ are the
generalized constitutive quantities of the flow, whereas
$\bm{\Delta}_{3}^{*}$ and $\bm{\Delta}_{4,2}^{*}$ are their associated
higher-order fluxes. Specifically, the viscous-stress flux
$\bm{\Delta}_{3}^{*}$ provides the higher-order closure associated with the
transport of $\bm{\Delta}_{2}^{*}$, whereas the flux associated with the
heat flux, $\bm{\Delta}_{4,2}^{*}$, provides the corresponding closure for
$\bm{\Delta}_{3,1}^{*}$. Equations~(\ref{e26}) and~(\ref{e27}) show that
these higher-order fluxes enter the evolution equations for
$\bm{\Delta}_{2}^{*}$ and $\bm{\Delta}_{3,1}^{*}$ and therefore control
their transport and relaxation dynamics.

The CE expansion is therefore used to derive the third-order
contributions to $\bm{\Delta}_{2}^{*}$ and $\bm{\Delta}_{3,1}^{*}$,
together with the second-order contributions to $\bm{\Delta}_{3}^{*}$ and
$\bm{\Delta}_{4,2}^{*}$. Taking moments of Eq.~(\ref{e7}) with the collision
invariants $(1,\mathbf{v}_i,(v_i^2+\eta_i^2)/2)$ gives the first-order
relations between $\partial_{t_1}$ and $\bm{\nabla}_1$:
\begin{equation}  \label{e31}
\partial_{t_1} \rho = -\bm{\nabla}_1 \cdot (\rho \mathbf{u}),
\end{equation}
\begin{equation}  \label{e32}
\partial_{t_1} \mathbf{u} = -R \bm{\nabla}_1 T
- \frac{R T}{\rho} \bm{\nabla}_1 \rho
- \mathbf{u} \cdot \bm{\nabla}_1 \mathbf{u},
\end{equation}
\begin{equation}  \label{e33}
\partial_{t_1} T = -\mathbf{u} \cdot \bm{\nabla}_1 T
- \frac{2 T}{n+3} \bm{\nabla}_1 \cdot \mathbf{u}.
\end{equation}
Substitution of these relations gives the first-order terms
$\bm{\Delta}_3^{*(1)}$ and $\bm{\Delta}_{4,2}^{*(1)}$, whose explicit forms
are listed in Table~\ref{TableIII}.

Applying the same procedure to Eq.~(\ref{e8}) gives the second-order
time-scale relations,
\begin{equation}  \label{e34}
\partial_{t_2} \rho=0,
\end{equation}
\begin{equation}  \label{e35}
\rho \partial_{t_2} \mathbf{u}
=-\bm{\nabla}_1 \cdot \bm{\Delta}_2^{(1)},
\end{equation}
\begin{equation}  \label{e36}
\partial_{t_2}\left[\frac{(n+3)}{2} \rho T
+\frac{1}{2} \rho u^2\right]
=-\bm{\nabla}_1 \cdot \bm{\Delta}_{3,1}^{(1)} .
\end{equation}
These relations yield the second-order terms $\bm{\Delta}_3^{*(2)}$ and
$\bm{\Delta}_{4,2}^{*(2)}$, whose explicit forms are listed in
Tables~\ref{TableIV}--\ref{TableV}. The second-order expressions for the
associated higher-order fluxes are therefore
\begin{equation}  \label{e37}
\bm{\Delta}_3^*
=\bm{\Delta}_3^{*(1)}+\bm{\Delta}_3^{*(2)},
\end{equation}
\begin{equation}  \label{e38}
\bm{\Delta}_{4,2}^*
=\bm{\Delta}_{4,2}^{*(1)}+\bm{\Delta}_{4,2}^{*(2)}.
\end{equation}

The third-order time-scale relations follow from Eq.~(\ref{e9}):
\begin{equation}  \label{e39}
\partial_{t_3} \rho=0,
\end{equation}
\begin{equation}  \label{e40}
\rho \partial_{t_3} \mathbf{u}
=-\bm{\nabla}_1 \cdot \bm{\Delta}_2^{(2)},
\end{equation}
\begin{equation}  \label{e41}
\partial_{t_3}\left[\frac{(n+3)}{2} \rho T
+\frac{1}{2} \rho u^2\right]
=-\bm{\nabla}_1 \cdot \bm{\Delta}_{3,1}^{(2)} .
\end{equation}
Substitution of these relations gives the third-order contributions to the
viscous stress and heat flux,
\begin{equation}  \label{e42}
\bm{\Delta}_2^{*(3)}
= \sum_i f_i^{(3)} \mathbf{v}_i^* \mathbf{v}_i^*
= -(\bm{\sigma}_{\text{Super-Burnett}}
-\bm{\sigma}_{\text{Burnett}}),
\end{equation}
\begin{equation}  \label{e43}
\bm{\Delta}_{3,1}^{*(3)}
= \sum_i f_i^{(3)} \frac{v_i^{*2}+\eta_i^2}{2}\mathbf{v}_i^*
= -(\mathbf{j}_{q,\text{Super-Burnett}}
-\mathbf{j}_{q,\text{Burnett}}).
\end{equation}
These terms represent the super-Burnett corrections to the Burnett-level
constitutive relations under the present sign convention.

To simplify the mechanism analysis, we focus on the $x$-direction
components of the viscous stress and heat flux. The model remains fully
three-dimensional, and the other components can be obtained in the same
way. The explicit third-order expressions for $\Delta_{2xx}^{*(3)}$ and
$\Delta_{3,1x}^{*(3)}$ are listed in Tables~\ref{TableVI}--\ref{TableVII}.
The resulting third-order constitutive expansions are
\begin{equation}  \label{e44}
\bm{\Delta}_2^*
= \bm{\Delta}_2^{*(1)}
+\bm{\Delta}_2^{*(2)}
+\bm{\Delta}_2^{*(3)},
\end{equation}
\begin{equation}  \label{e45}
\bm{\Delta}_{3,1}^*
= \bm{\Delta}_{3,1}^{*(1)}
+\bm{\Delta}_{3,1}^{*(2)}
+\bm{\Delta}_{3,1}^{*(3)}.
\end{equation}
Equations~(\ref{e37})--(\ref{e38}) and~(\ref{e44})--(\ref{e45}) define the
TNE measures used below to analyse multigradient coupling in high-order
nonequilibrium responses.

\section{Model verification}
\label{Numerical simulations}

This section implements the kinetic moment constraints derived above through
D3V91 discrete velocity sets. The resulting model is then tested using
shock-tube benchmarks to assess its ability to reproduce the macroscopic
wave structures of high-speed compressible flows. Finally, the influence of
the spatial discretization scheme on high-order TNE quantities is examined,
because these quantities are more sensitive to numerical smoothness and
resolution than conserved macroscopic fields.

\subsection{Phase-space discretization scheme}

Phase-space discretization is fundamental to both the formulation and the
numerical implementation of the DBM. The discrete velocity set must preserve
the target kinetic moment relations using a finite number of velocity
points. Its structure therefore affects both the numerical behaviour of the
discrete Boltzmann equation and the ability of the model to resolve
high-order TNE effects. Based on the moment constraints derived in the
previous section, we construct a first D3V91 discrete velocity set, denoted
D3V91(I), as shown in Fig.~\ref{Fig01}. The corresponding discrete velocity
values are listed in Table~\ref{TableI} of the Appendix.

\begin{figure}
\centering
\includegraphics[width=0.6\textwidth]{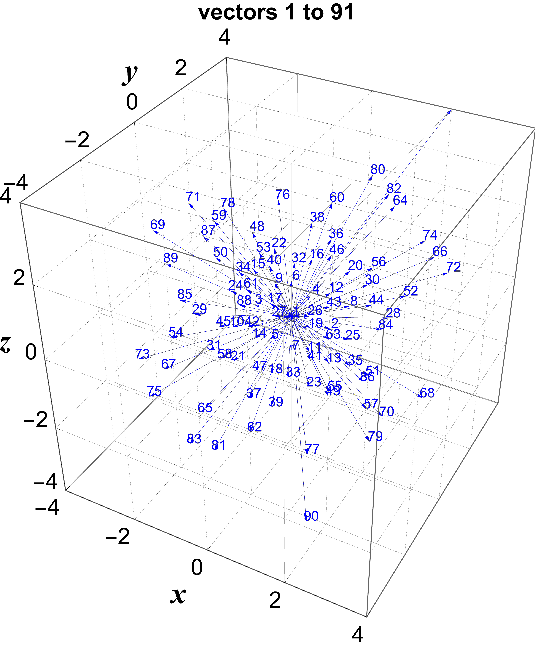}
\caption{Schematic diagram of the D3V91(I) discrete velocity set.}
\label{Fig01}
\end{figure}

\begin{figure}
\centering
\includegraphics[width=0.8\textwidth]{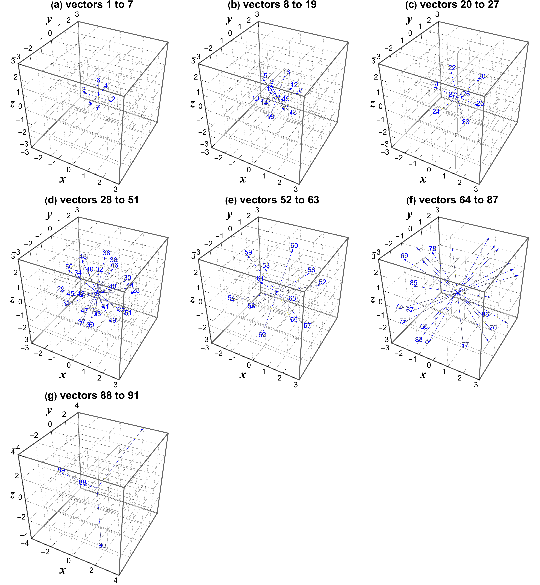}
\caption{Decomposition of the D3V91(I) discrete velocity set.}
\label{Fig02}
\end{figure}

Figure~\ref{Fig02} shows the decomposition of D3V91(I). The first velocity
is the zero velocity. Velocities 2--7 are face-centred velocities;
velocities 8--19 are second-diagonal velocities; velocities 20--27 are
cube-vertex velocities; velocities 28--51 are third-diagonal velocities;
velocities 52--63 are also second-diagonal velocities; velocities 64--87 are
higher-order mixed-diagonal velocities; and the last four velocities are
special velocities with reduced symmetry, introduced to make the moment
matrix invertible.
The internal degree parameter is prescribed as
\begin{equation*}
\eta _{i}=%
\begin{cases}
10\eta _{0} & i=1, \\
\eta _{0} & 2\leq i\leq 7,\,i=15,\,19,\,20,\,26, \\
& 35\leq i\leq 40,\,52\leq i\leq 53,\,60,\,64,\,65, \\
0 & \text{otherwise},
\end{cases}%
\end{equation*}
where $c$ and $\eta _{0}$ are two free parameters introduced to ensure the
existence of the inverse matrix $\mathbf{C}^{-1}$.

\begin{figure}
\centering
\includegraphics[width=0.6\textwidth]{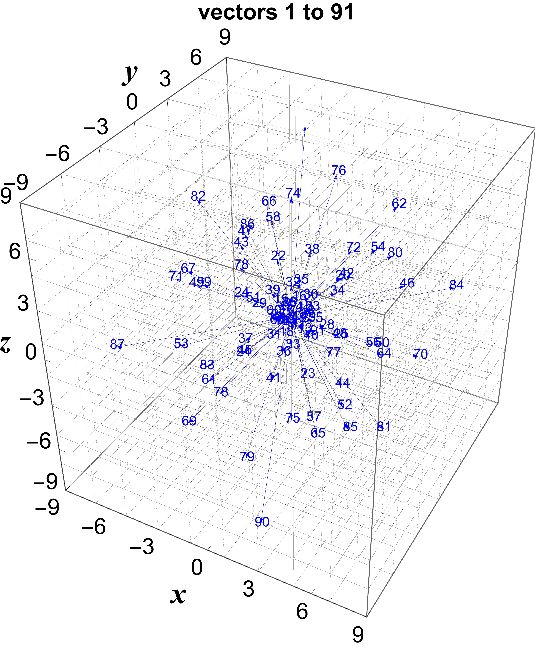}
\caption{Schematic diagram of the D3V91(II) discrete velocity set.}
\label{Fig03}
\end{figure}

\begin{figure}
\centering
\includegraphics[width=0.8\textwidth]{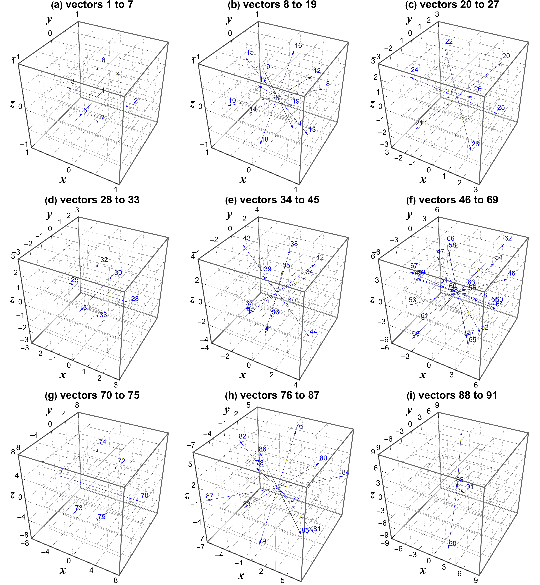}
\caption{Decomposition of the D3V91(II) discrete velocity set.}
\label{Fig04}
\end{figure}

The D3V91(I) set is mainly used to describe
$\bm{\Delta}_3^{*(1)+(2)}$ and $\bm{\Delta}_{4,2}^{*(1)+(2)}$. To improve
the representation of $\bm{\Delta}_2^{*(1)+(2)+(3)}$ and
$\bm{\Delta}_{3,1}^{*(1)+(2)+(3)}$, we further modify the velocity set and
construct a second realization, denoted D3V91(II). As shown in
Figs.~\ref{Fig03} and~\ref{Fig04}, D3V91(II) has two main features. First,
the magnitude of the discrete velocities is extended from 0.7 to 12.9,
covering a wider velocity range. Second, additional discrete velocity
directions are introduced while the overall symmetry of the velocity set is
preserved. These modifications in velocity magnitude and direction improve
the representation of distribution functions in high-speed compressible
flows with strong discontinuities and enhance numerical robustness in
strongly nonequilibrium regimes \citep{33Wu2025POF}. The detailed discrete
velocity values of this model are listed in Table~\ref{TableII} of the
Appendix.
In D3V91(II), $\eta _{i}$ is given by
\begin{equation*}
\eta _{i}=%
\begin{cases}
10\eta _{0} & i=1, \\
\eta _{0} & 2\leq i\leq 7,\,i=11,\,22,\,26,\,32,\,40,\,43, \\
& 51,\,58,\,63,\,68,\,75,\,83,\,84,\,86, \\
0 & \text{otherwise}.
\end{cases}%
\end{equation*}

The DBM specifies the moment constraints required to represent the target
physical system, but it does not impose a unique discretization in time,
space or velocity space. The D3V91 discrete velocity sets used here are
therefore feasible realizations that satisfy the present moment constraints,
rather than uniquely optimal choices in a general sense.

We next use one-dimensional shock-tube benchmarks to test whether the model
reproduces the macroscopic structures generated by strong gradients. Time
integration uses a second-order implicit--explicit Runge--Kutta
finite-difference scheme \citep{52Ascher1997ANM}. The spatial derivative is
discretized using either a fifth-order weighted essentially non-oscillatory
(5th WENO) scheme \citep{53Liu1994JOCP} or a second-order non-oscillatory,
no-parameter dissipation (2nd NND) scheme \citep{54Zhang1988AAS}.

\textbf{Sod shock tube}

The initial condition for the Sod shock tube problem \citep{55Sod1978JOCP}
is
\begin{equation}  \label{ee1}
\left\{
\begin{array}{l}
\left( \rho, T, u_x, u_y, u_z \right)|_L = \left( 1.0,1.0,0.0,0.0,0.0
\right), \\
\left( \rho, T, u_x, u_y, u_z \right)|_R = \left(
0.125,0.8,0.0,0.0,0.0\right).
\end{array}
\right.
\end{equation}
For D3V91(II), the parameters are $\tau=6 \times 10^{-5}$,
$n=0$, $c=2$, $\eta_0=10$, $\Delta x = \Delta y = \Delta z =
3 \times 10^{-3}$ and $\Delta t = 10^{-4}$.

\begin{figure}
\centering
\includegraphics[width=0.90\textwidth]{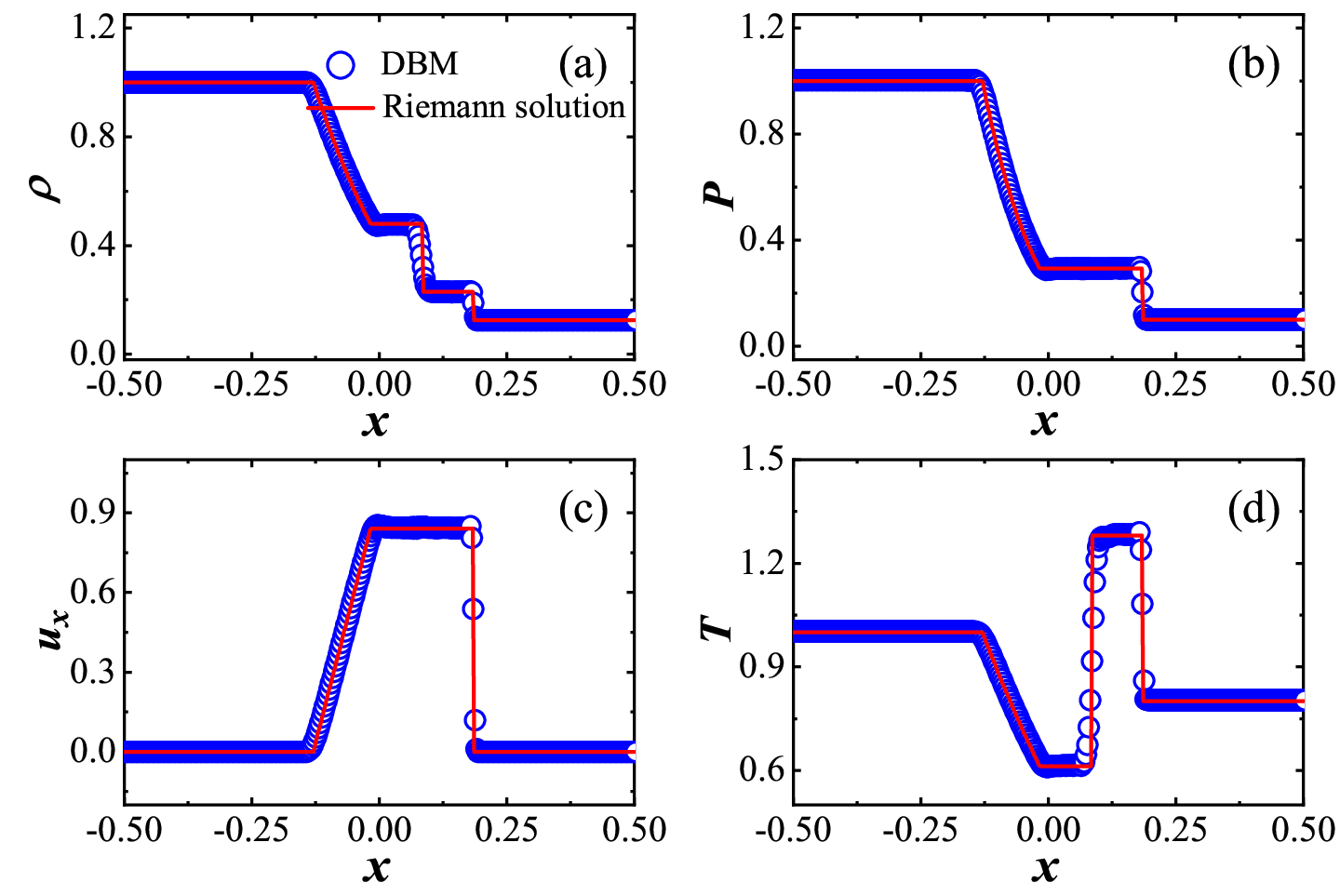}
\caption{Comparison between the DBM simulation and the Riemann solution for
the Sod shock tube at $t = 0.1$: (a) density, (b) pressure, (c) velocity and
(d) temperature.}
\label{Fig05}
\end{figure}

Figure~\ref{Fig05} compares the DBM results with the Riemann solution for
the Sod shock tube at $t = 0.1$. The density, pressure, velocity and
temperature profiles are consistent with the reference solution. The model
reproduces the main Riemann wave structures, including the shock wave,
contact discontinuity and rarefaction wave. This benchmark tests the
macroscopic consistency of D3V91(II) in a one-dimensional compressible
Riemann problem. Similar results are obtained with D3V91(I).

\textbf{Lax shock tube}

The initial condition for the Lax shock tube problem is
\begin{equation}  \label{ee2}
\left\{
\begin{array}{l}
\left( \rho, T, u_x, u_y, u_z \right)|_L = \left( 0.445,7.928,0.698,0.0,0.0
\right), \\
\left( \rho, T, u_x, u_y, u_z \right)|_R = \left(
0.50,1.142,0.0,0.0,0.0\right),
\end{array}
\right.
\end{equation}
where $\tau=5 \times 10^{-5}$, $n=0$, $c=1.5$, $\eta_0=7.6$,
$\Delta x = \Delta y = \Delta z = 3 \times 10^{-3}$ and
$\Delta t = 5 \times 10^{-4}$.

\begin{figure}
\centering
\includegraphics[width=0.90\textwidth]{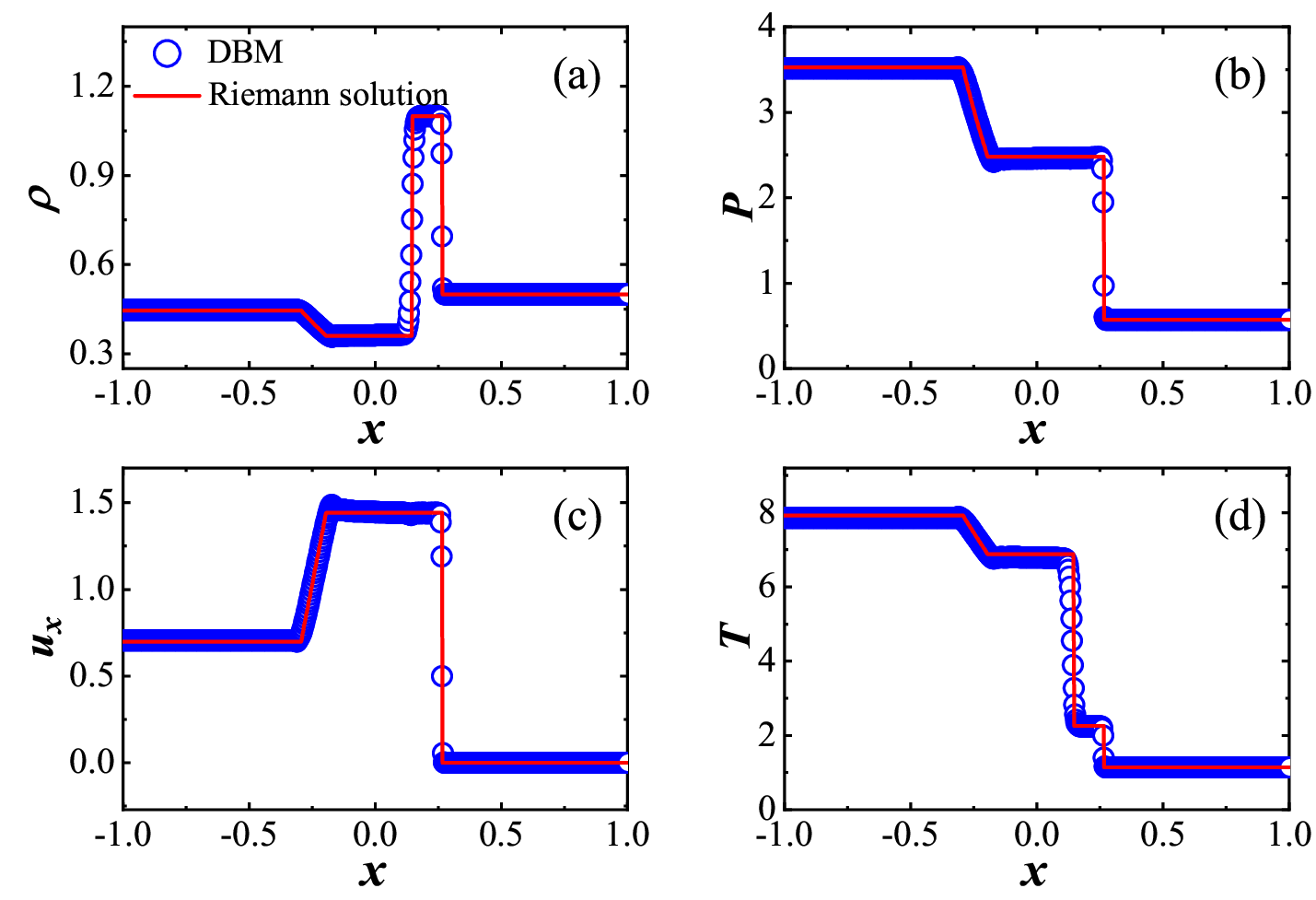}
\caption{Comparison between the DBM simulation and the Riemann solution for
the Lax shock tube at $t = 0.1$: (a) density, (b) pressure, (c) velocity and
(d) temperature.}
\label{Fig06}
\end{figure}

Figure~\ref{Fig06} compares the DBM results with the Riemann solution for
the Lax shock tube. Compared with the Sod case, this problem contains a
larger temperature contrast and a non-zero left-state velocity. The density,
pressure, velocity and temperature profiles remain consistent with the
reference solution, indicating that the model can reproduce the main
macroscopic wave structures under stronger temperature and velocity
gradients.

\subsection{Influence of numerical schemes on the description of TNE effects}

To characterize the nonequilibrium state, we introduce an 11-component
nonequilibrium intensity vector,
$\mathbf{S}_\text{TNE}$ = ($\tau$, $\rho$, $T$, $\mathbf{u}$,
$\bm{\nabla} \rho$, $\bm{\nabla} T$, $\bm{\nabla} \mathbf{u}$,
$\bm{\Delta}_2^{*(1)+(2)+(3)}$,
$\bm{\Delta}_{3,1}^{*(1)+(2)+(3)}$, $\bm{\Delta}_3^{*(1)+(2)}$,
$\bm{\Delta}_{4,2}^{*(1)+(2)}$). This vector collects the relaxation
parameter, macroscopic fields, gradients and non-conserved kinetic moments
that determine the TNE response.

To assess whether the third-order DBM resolves nonequilibrium effects at
different asymptotic orders, we perform a series of fluid-collision tests.
By varying the initial states and physical parameters, we generate TNE
responses dominated by different-order contributions. The initial conditions
are
\begin{equation}  \label{ee3}
\rho(x, y, z)=\frac{\rho_L+\rho_R}{2}-\frac{\rho_L-\rho_R}{2} \tanh \left(%
\frac{x-N_x \Delta x / 2}{L_\rho}\right),
\end{equation}
\begin{equation}  \label{ee4}
T(x, y, z)=\frac{T_L+T_R}{2}-\frac{T_L-T_R}{2} \tanh \left(\frac{x-N_x
\Delta x / 2}{L_T}\right),
\end{equation}
\begin{equation}  \label{ee5}
u_x(x, y, z)=-u_0 \tanh \left(\frac{x-N_x \Delta x / 2}{L_u}\right).
\end{equation}
Here $L_\rho$, $L_u$ and $L_T$ denote the widths of the density, velocity
and temperature transition layers, respectively. The quantities $\rho_L$ and
$\rho_R$ are the left- and right-state densities, and $T_L$ and $T_R$ are
the corresponding temperatures. The transverse velocities are set to
$u_y=0$ and $u_z=0$. The computational domain is a cuboid of size
$1.5 \times 0.006 \times 0.006$, discretised by a uniform grid of
$1000 \times 4 \times 4$.

Numerical schemes play a critical role in realizing the resolving capability
of a high-order DBM. Although the moment constraints determine which TNE
quantities can be represented, the numerical scheme must provide
sufficiently smooth and accurate macroscopic fields and gradients for these
quantities to be evaluated reliably.

\begin{figure}
  \centering
\includegraphics[width=0.9\textwidth]{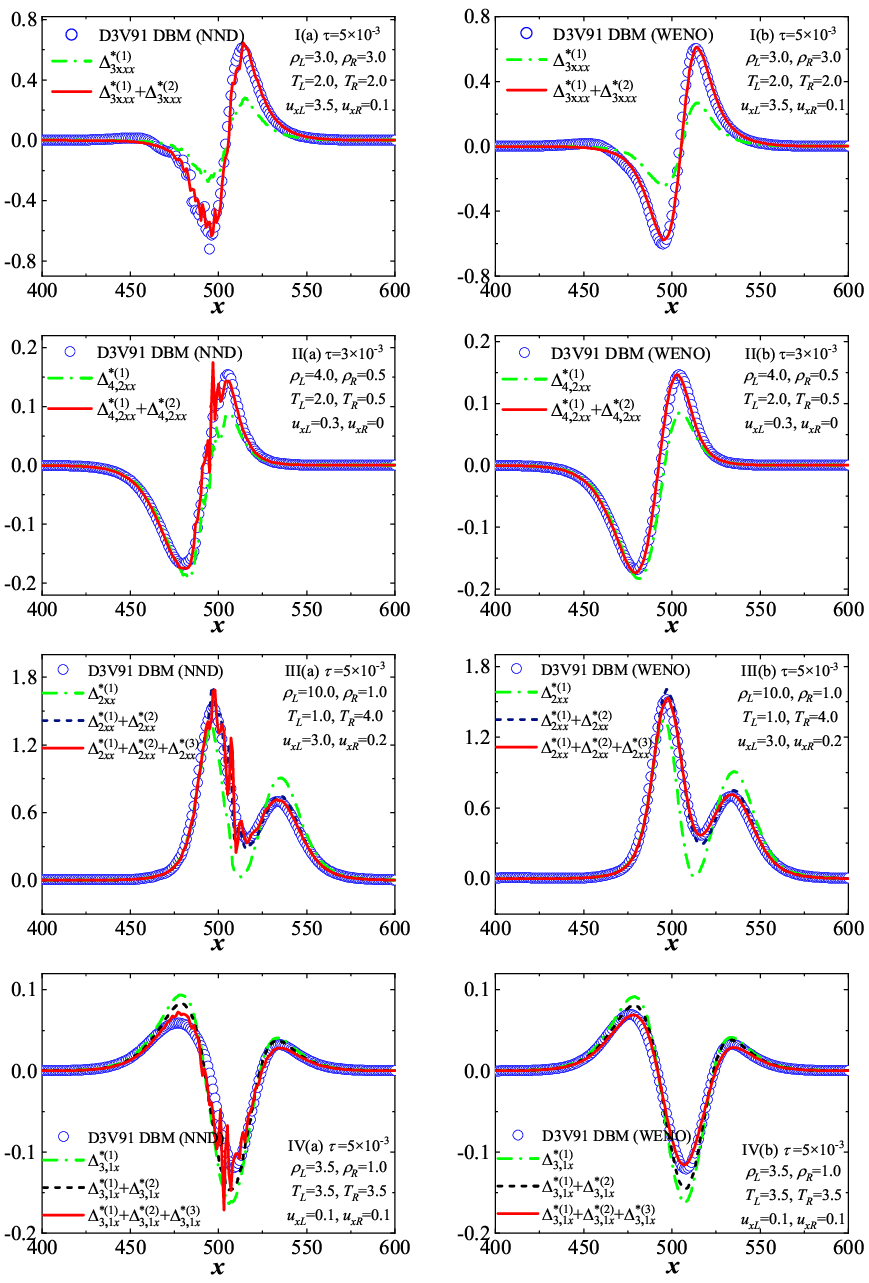}
\caption{Distributions of the TNE quantities computed using the
2nd NND scheme (left column) and the 5th WENO scheme
(right column).
Blue open circles denote the DBM results. The green dash-dotted, black
dashed and red solid lines denote the first-, second- and third-order
analytical solutions, respectively.}
\label{Fig07}
\end{figure}

Figure~\ref{Fig07} compares the TNE quantities
$\bm{\Delta}_3^*$, $\bm{\Delta}_{4,2}^*$, $\bm{\Delta}_2^*$ and
$\bm{\Delta}_{3,1}^*$ computed using the 2nd NND and 5th WENO schemes under
the same initial conditions. Blue open circles denote the DBM results.
For ${\Delta}_{3xxx}^*$ and ${\Delta}_{4,2xx}^*$, the green dash-dotted and red solid lines denote the first- and second-order analytical solutions, respectively.
For ${\Delta}_{2xx}^*$ and ${\Delta}_{3,1x}^*$, the green dash-dotted, dark-blue dashed and red solid lines denote the first-, second- and third-order analytical solutions, respectively.

When the 2nd NND scheme is employed, clear discrepancies remain between the
numerical profiles and the high-order analytical expressions, even with the
high-order DBM. Oscillations also appear in the evaluated high-order TNE profiles.
This indicates that a second-order scheme does not provide sufficiently smooth or accurate macroscopic fields and gradients for reliable evaluation of high-order TNE quantities.

By contrast, the fifth-order WENO scheme gives results that are consistent with the high-order analytical expressions. Its adaptive weighting suppresses nonphysical oscillations and reduces excessive numerical dissipation, yielding smoother macroscopic fields and gradients.
The results also show that different TNE quantities have distinct multiscale characteristics. Under the present conditions, second-order contributions are sufficient to describe $\bm{\Delta}_3^*$ and $\bm{\Delta}_{4,2}^*$, whereas $\bm{\Delta}_2^*$ and $\bm{\Delta}_{3,1}^*$ require third-order contributions.
These comparisons indicate that, as the TNE intensity increases and the dominant asymptotic order changes, the numerical scheme must resolve the relevant physical scales to realize the moment-closure capability of the model. This provides the numerical basis for the
multigradient coupling analysis in the following sections.

\section{Capability to resolve nonequilibrium effects}
\label{TNE}

This section examines the ability of the D3V91 super-Burnett-level DBM to
resolve the high-order TNE quantities considered here, namely
$\bm{\Delta}_3^*$, $\bm{\Delta}_{4,2}^*$, $\bm{\Delta}_2^*$ and
$\bm{\Delta}_{3,1}^*$. The analysis has two aims. First, we test whether
the D3V91 model, which satisfies the super-Burnett-level moment constraints,
reproduces constitutive relations beyond the Burnett level. Second, we use
these TNE quantities to determine how leading gradients, secondary gradients
and transition-layer widths jointly regulate nonequilibrium responses. This
section therefore links kinetic moment closure to the multigradient-coupling
mechanisms analysed below.

\subsection{Capability to resolve $\bm{\Delta}_3^*$}

To test the ability of the model to resolve $\bm{\Delta}_3^*$, we consider
three cases with increasing TNE intensity. The initial jumps and
transition-layer widths are prescribed as follows:

(i) $\rho_L=0.1, \rho_R=0.1, T_L=2, T_R=0.1, u_{xL}=0, u_{xR}=0,
L_\rho=L_u=L_T=20, \tau = 2 \times 10^{-3}$;

(ii) $\rho_L=1, \rho_R=1, T_L=1, T_R=1, u_{xL}=2, u_{xR}=0.1,
L_\rho=L_u=L_T=20, \tau = 2 \times 10^{-3}$;

(iii) $\rho_L=3, \rho_R=3, T_L=3, T_R=3, u_{xL}=3.5, u_{xR}=0.1,
L_\rho=L_u=L_T=10, \tau = 3 \times 10^{-3}$.

\begin{figure}
 \centering
\includegraphics[width=0.9\textwidth]{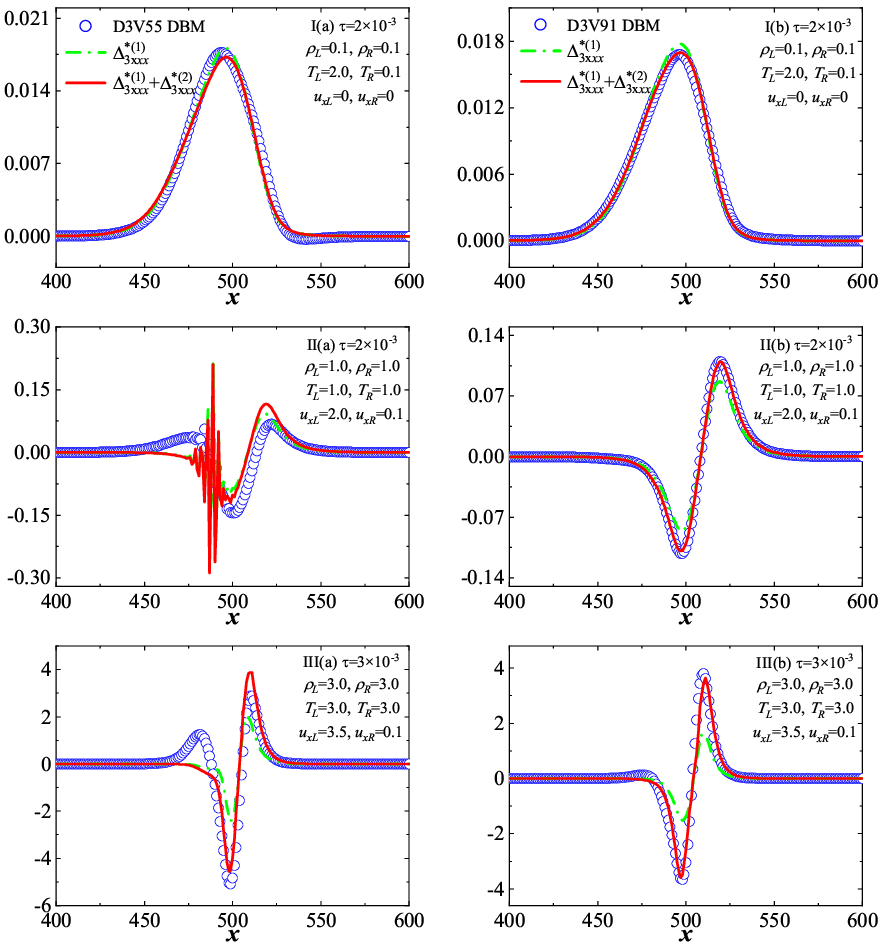}
\caption{Distributions of ${\Delta}_{3xxx}^*$ for the weak (I), moderate
(II), and strong (III) cases computed using the D3V55 model (left column)
and the D3V91 model (right column).
Blue open circles denote the DBM results. The green dash-dotted and red solid lines denote the first- and second-order analytical solutions, respectively.
The corresponding times for cases I--III are $t$=0.012, 0.012, and 0.014, respectively.}
\label{Fig08}
\end{figure}

Figure~\ref{Fig08} compares ${\Delta}_{3xxx}^*$ obtained with the
Burnett-level D3V55 model and the super-Burnett-level D3V91 model.
In case I, only a temperature gradient is imposed initially. The constitutive
expression for $\Delta_{3xxx}^*$ shows that the first-order term
$\Delta_{3xxx}^{*(1)}$ is proportional to $\partial_x T$ and therefore
dominates the TNE response. At the initial stage, the velocity-curvature term
and the velocity--temperature-gradient coupling terms in the second-order
contribution vanish (see Appendix table~\ref{TableV}, $S_{37}$--$S_{39}$).
During the early evolution, the imposed temperature gradient has not yet
generated an appreciable velocity gradient, so $\Delta_{3xxx}^{*(2)}$
remains weak. Figures~\ref{Fig08}I(a,b) show that the first- and
second-order analytical profiles almost coincide. The maximum relative TNE
intensity,
$R_{\text{TNE}} =
|\Delta_{3xxx}^{*(2)} / \Delta_{3xxx}^{*(1)}|$
\citep{50Gan2022JFM}, is only 0.046 at $x \approx 496$, confirming that the
second-order contribution is negligible. Under these conditions, both D3V55
and D3V91 resolve the dominant first-order response. The lower-order model,
however, shows noticeable deviations near the TNE peak.

In case II, the temperature is initially uniform, so
$\Delta_{3xxx}^{*(1)}=0$ at the initial time. The second-order contribution
$\Delta_{3xxx}^{*(2)}$ is then driven by the velocity gradient through the
velocity-curvature and velocity-gradient-product terms (see Appendix
table~\ref{TableV}). At $t=0.012$, however, $\Delta_{3xxx}^{*(1)}$ becomes
dominant, while $R_{\text{TNE}}$ remains 0.31. This suggests that
the second-order contribution modifies the temperature-gradient-driven
first-order response through viscous-stress work, but subsequently loses
its dominant role.
Although the first-order contribution dominates at this
time, the D3V55 model does not contain the moment constraints required to
represent the formation of $\Delta_{3xxx}^{*(2)}$ and its feedback on the
temperature gradient. As a result, its numerical solution deviates from the
constitutive description. The oscillatory behaviour in the evaluated
higher-order profiles is associated with discontinuities in the macroscopic
fields.

In case III, the density, temperature and relaxation time are increased, the
velocity gradient is strengthened, and the interface width is reduced. The
strong velocity gradient then makes the second-order contribution dominate
the TNE response. The relative TNE intensity reaches
$R_{\text{TNE}}\approx 1.28$ at $x \approx 511$, well above 0.5. Owing to
the larger macroscopic amplitudes and relaxation time, the TNE intensity is
approximately 40 times larger than that in case II. The D3V55 model lacks
the high-order non-conserved moment constraints required in this regime and therefore
deviates from the constitutive description. By contrast, the D3V91 model
satisfies the moment relations required to describe
$\Delta_{3xxx}^{*(1)}$ and $\Delta_{3xxx}^{*(2)}$, and its numerical results
agree closely with the second-order analytical solution. These comparisons show that resolving $\bm{\Delta}_3^*$ requires super-Burnett-level moment constraints when the second-order contribution becomes comparable to, or larger than, the first-order contribution.

\subsection{Nonequilibrium phase diagram of $\bm{\Delta}_3^*$}

High-order TNE quantities are generally controlled by coupled macroscopic
gradients rather than by a single driving gradient. Each quantity has a
leading-gradient dependence associated with its lowest-order constitutive
contribution, whereas secondary gradients, gradient products, curvature
terms and transition-layer widths modify the higher-order contributions. A
complete description of high-order TNE therefore requires both the
leading-gradient response and the secondary-gradient corrections. We first
test whether the D3V91 model resolves the target TNE quantities and then
analyse how different gradients and transition-layer widths regulate their
multigradient-coupled responses.

For $\bm{\Delta}_3^*$, the temperature gradient $\partial_x T$ is the
leading driver of $\Delta_{3xxx}^{*(1)}$, whereas secondary gradients and
gradient-coupling terms contribute to $\Delta_{3xxx}^{*(2)}$. These include
second-order velocity derivatives ($\partial^2_x u_x$,
$\partial^2_{xy} u_y$, $\partial^2_{xz} u_z$) and
temperature--velocity coupling terms
($\partial_y T \partial_y u_x$, $\partial_z T \partial_z u_x$). We therefore
examine the velocity-gradient and density-gradient effects on
$\bm{\Delta}_3^*$ to clarify how secondary gradients modify its leading
temperature-gradient response.

\begin{figure}
 \centering
\includegraphics[width=1\textwidth]{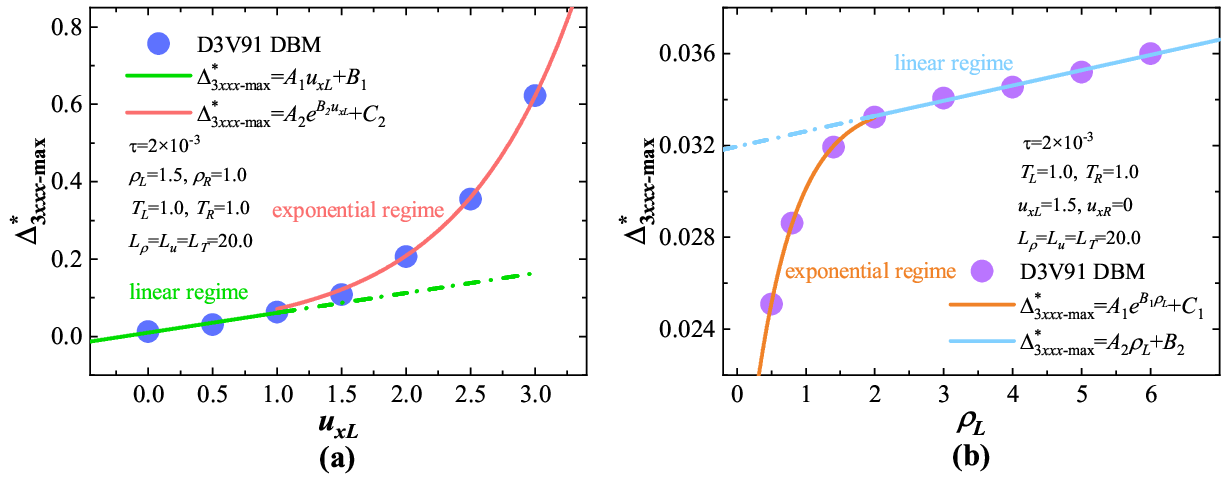}
\caption{Effects of the velocity gradient (a) and density gradient (b) on the
nonequilibrium quantity $\bm{\Delta}_3^*$: linear and exponential responses, where for (a): $A_1 = 0.051$, $B_1 = 0.010$, $A_2 = 0.023$, $B_2 = 1.102$, $C_2 = 0.003$; for (b): $A_1 = 0.011$, $B_1 = 0.713$, $C_1 = 0.016$, $A_2 = 0.001$, $B_2 = 0.032$.}
\label{Fig09}
\end{figure}

Figure~\ref{Fig09}(a) shows the velocity-gradient effect on
$\Delta_{3xxx\text{-max}}^{*}$. The right-state velocity is fixed at $-0.5$,
while the left-state velocity is increased. As $\partial_x u_x$ increases,
$\Delta_{3xxx\text{-max}}^{*}$ exhibits two-stage growth behaviour:

(i) Linear growth stage (0 < $u_{xL}$ < 1): $\Delta_{3xxx-\max}^* = A_1
u_{xL} + B_1$.
In the absence of an initial temperature gradient, and for weak velocity
gradients, the relevant terms in $\Delta_{3xxx}^{*(2)}$ are approximately
linear in $\partial_x u_x$, such as the contribution proportional to
$(5+3n)\partial_x u_x$. Nonlinear contributions from second-order velocity
derivatives, squared velocity-gradient terms and temperature--velocity
coupling terms remain weak. The TNE intensity therefore increases
approximately linearly with $\partial_x u_x$.

(ii) Exponential growth stage ($u_{xL}$ > 1): $\Delta_{3xxx-\max}^* = A_2
e^{B_2 u_{xL}} + C_2$.
As the velocity gradient increases, second-order velocity derivatives and
temperature--velocity coupling terms become more pronounced. These
additional transport channels amplify the TNE response, producing an
approximately exponential increase in $\Delta_{3xxx}^{*}$.

Thus, although the velocity gradient is not the leading driver of
$\Delta_{3xxx}^{*(1)}$, it strongly modulates $\Delta_{3xxx}^{*}$ through
second-order coupling channels. This sensitivity indicates that
$\Delta_{3xxx}^{*}$ contains both energy-transport information and
momentum-gradient-induced coupling effects.

Figure~\ref{Fig09}(b) shows the density-gradient effect on
$\Delta_{3xxx\text{-max}}^{*}$, with the right-state density fixed at
$\rho_R=0.5$. As $\rho_L$ increases, the absolute density difference increases and
$\Delta_{3xxx\text{-max}}^{*}$ exhibits two response regimes.
This case provides a reference for the density-gradient
responses discussed below. The relevant
driving measure, however, is the relative density variation,
$\bm{\nabla}\rho/\rho$, rather than the absolute density gradient alone.

(i) Exponential growth stage ($0.5<\rho_L<2$):
$ \Delta_{3xxx-\max}^* = A_1 e^{B_1 \rho_L} + C_1 .$
In this low-density regime, even a modest absolute density variation can
produce a large relative density gradient. The enhanced relative variation
strengthens compressibility effects and activates nonlinear transport
channels, including higher-order velocity and temperature derivatives and
their gradient-product couplings. These coupled contributions amplify
$\Delta_{3xxx}^{*}$ and produce an approximately exponential increase in
$\Delta_{3xxx\text{-max}}^{*}$.


(ii) Linear growth stage ($\rho_L>2$): $\Delta_{3xxx-\max}^* = A_2 \rho_L + B_2$.
As the background density increases, the relative density variation weakens
even though the absolute density difference continues to grow. The
compressibility-driven nonlinear couplings are therefore reduced, and the
response becomes governed mainly by leading linear driving terms.
Consequently, $\Delta_{3xxx\text{-max}}^{*}$ varies approximately linearly
with $\rho_L$.


\subsection{Capability to resolve $\bm{\Delta}_{4,2}^*$}

To test the ability of the model to resolve $\bm{\Delta}_{4,2}^*$, we
consider three cases with increasing TNE intensity:

(i) $\rho_L=0.5, \rho_R=0.5, T_L=0.5, T_R=0.5, u_{xL}=1, u_{xR}=0,
L_\rho=L_u=L_T=20, \tau = 1 \times 10^{-3}$;

(ii) $\rho_L=2, \rho_R=0.5, T_L=2, T_R=0.5, u_{xL}=0.5, u_{xR}=0,
L_\rho=L_u=L_T=20, \tau = 2 \times 10^{-3}$;

(iii) $\rho_L=3, \rho_R=1, T_L=3, T_R=1, u_{xL}=0, u_{xR}=0,
L_\rho=L_u=L_T=8, \tau = 3 \times 10^{-3}$.

\begin{figure}
 \centering
\includegraphics[width=0.9\textwidth]{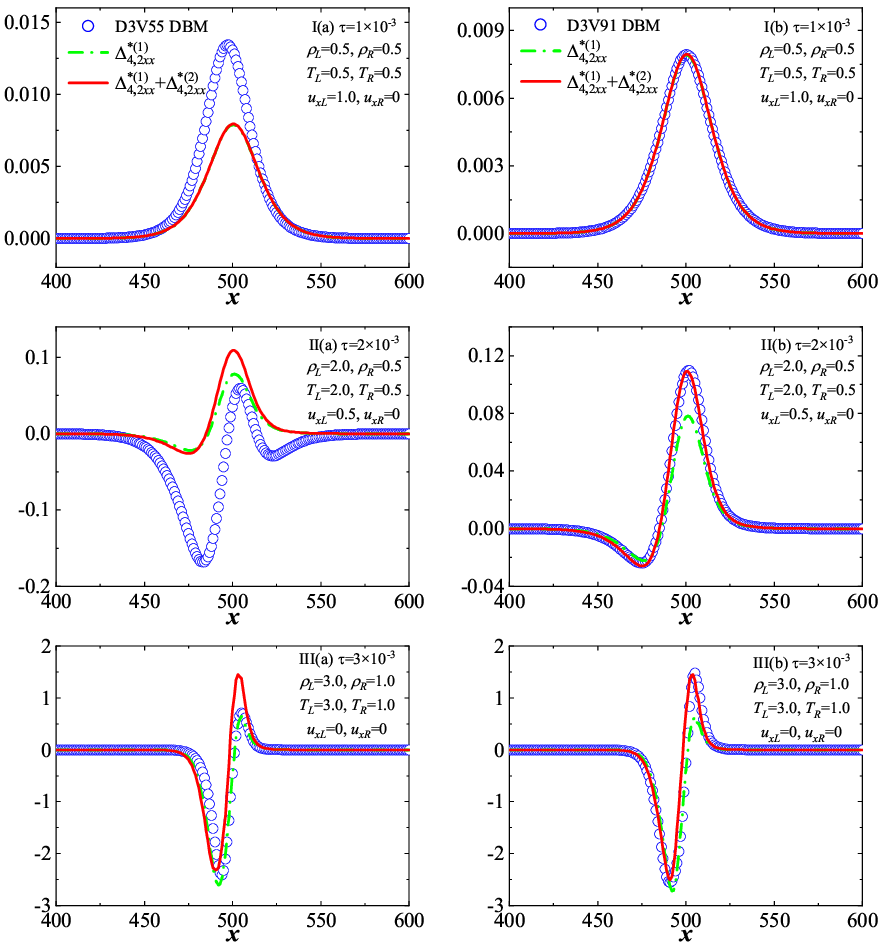}
\caption{Distributions of ${\Delta}_{4,2xx}^*$ for the weak (I), moderate
(II) and strong (III) cases computed using the D3V55 model (left column)
and the D3V91 model (right column). Blue open circles denote the DBM results. The green dash-dotted and red solid lines denote the first- and second-order analytical solutions, respectively. The corresponding times for cases I--III are $t=0.002$, $0.008$ and $0.0055$, respectively.}
\label{Fig10}
\end{figure}

Blue open circles denote the DBM results. The green dash-dotted and red solid lines denote the first- and second-order analytical expressions, respectively.

Figure~\ref{Fig10} compares $\Delta_{4,2xx}^{*}$ obtained with the
Burnett-level D3V55 model and the super-Burnett-level D3V91 model.

In case I, only a velocity gradient is imposed initially, while the density
and temperature are uniform. The constitutive expressions show that
$\Delta_{4,2xx}^{*(1)}$ is primarily driven by the velocity gradient,
whereas the density-, temperature- and gradient-coupling-related terms in
$\Delta_{4,2xx}^{*(2)}$ remain weak. The system therefore stays close to the
first-order TNE regime, and the first- and second-order analytical profiles
nearly coincide, as shown in Figs.~\ref{Fig10}I(a,b). Nevertheless, the
D3V55 model shows visible deviations near the TNE peak. This indicates that
$\Delta_{4,2xx}^{*}$ is sensitive to the moment-closure accuracy of the
discrete velocity model even when the relative TNE intensity is weak. The
D3V91 model, which satisfies the required higher-order moment constraints,
better resolves the TNE structure near the gradient extrema.

This comparison shows that the applicability of a discrete velocity model is
not determined solely by the magnitude of the TNE intensity. It also depends
on whether the discrete velocity set satisfies the moment relations required
by the target TNE quantity. Even in weakly nonequilibrium regimes,
incomplete moment closure can produce visible deviations near gradient
extrema, and these closure-induced deviations become more pronounced as the
TNE intensity increases.

In case II, density and temperature gradients are introduced, the relaxation
time is increased, and the velocity gradient is reduced. The first-order
contribution $\Delta_{4,2xx}^{*(1)}$ remains dominant, but nonlinear
second-order terms are enhanced. These terms include second-order density
and temperature derivatives, $\partial_x^2 \rho$ and $\partial_x^2 T$, and
squared-gradient terms such as $(\partial_x u_x)^2$, $(\partial_x T)^2$ and
$(\partial_x \rho)^2$. At this stage, $R_{\mathrm{TNE}}\approx 0.39$.
Figures~\ref{Fig10}II(a,b) show that D3V55 deviates from both the first- and
second-order analytical solutions, whereas D3V91 is consistent with the
second-order analytical solution. The Burnett-level closure is therefore
insufficient for representing $\Delta_{4,2xx}^{*(2)}$ in
multigradient-driven flows.

In case III, the density and temperature gradients are further increased,
the interface width is reduced, the relaxation time is increased, and the
velocity gradient is removed. The first-order contribution
$\Delta_{4,2xx}^{*(1)}$ is then reduced, while the terms $S_{66}$--$S_{67}$
and $S_{72}$--$S_{73}$ in $\Delta_{4,2xx}^{*(2)}$ become dominant. The
relative TNE intensity reaches $R_{\mathrm{TNE}}\approx 1.58$, indicating a
regime dominated by second-order TNE effects. The TNE intensity is about
15 times larger than that in case II. Under these stronger TNE conditions,
D3V91 remains consistent with the second-order analytical description,
whereas the lower-order closure is no longer sufficient. These comparisons
show that resolving $\bm{\Delta}_{4,2}^*$ requires the corresponding
super-Burnett-level moment constraints when the second-order contribution
dominates the TNE response.

\subsection{Nonequilibrium phase diagram of $\bm{\Delta}_{4,2}^*$}

Because $\bm{\Delta}_{4,2}^*$ is primarily governed by the velocity
gradient, we next examine how the density-gradient and temperature-gradient
effects modify its response.

\begin{figure}
\centering
\includegraphics[width=1\textwidth]{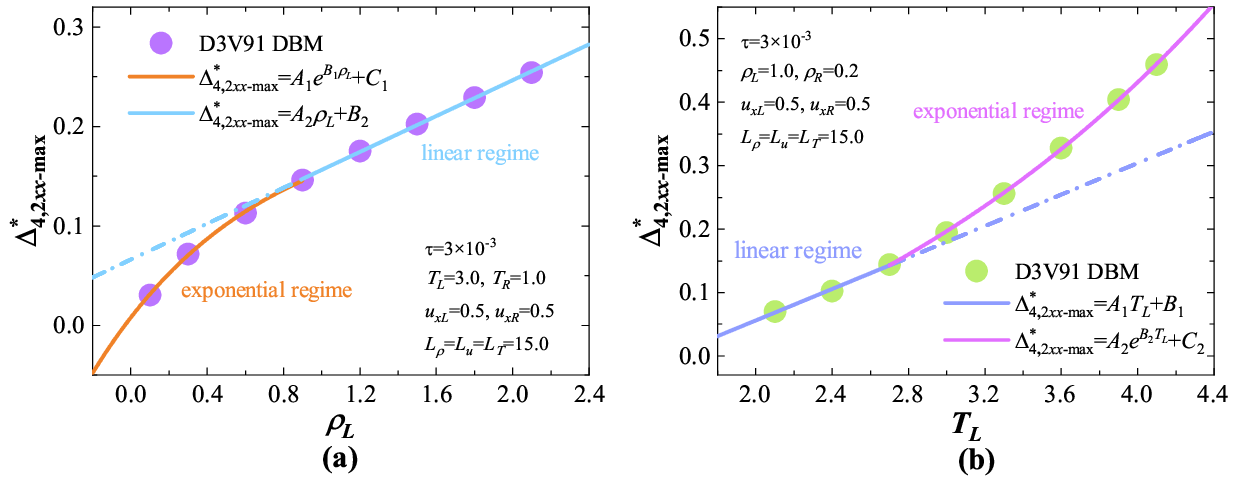}
\caption{Effects of the density gradient (a) and temperature gradient (b) on
the nonequilibrium quantity $\bm{\Delta}_{4,2}^*$: linear and exponential
responses, where for (a): $A_1 = 0.025$, $B_1 = 2.185$, $C_1 = 0.001$, $A_2 = 0.090$, $B_2 = 0.066$; for (b): $A_1 = 0.124$, $B_1 = -0.192$, $A_2 = 0.141$, $B_2 =0.404$, $C_2 = -0.275$.}
\label{Fig11}
\end{figure}

Figure~\ref{Fig11}(a) shows the density-gradient effect on
$\Delta_{4,2xx\text{-max}}^{*}$ in the absence of an initial velocity
gradient, with $\rho_R=0.2$. The quantity
$\Delta_{4,2xx\text{-max}}^{*}$ again exhibits two response regimes as $\rho_L$
increases.

(i) Exponential growth stage (0.1 < $\rho_L$ < 0.9): $\Delta_{4,2xx-%
\max}^*=A_1 e^{B_1\rho_L}+C_1$.
The exponential branch reflects the activation of second-order coupling
terms in $\Delta_{4,2xx}^{*(2)}$. Unlike $\Delta_{3xxx}^{*}$, this tensorial
heat-flux-related flux contains density-, temperature- and
velocity-gradient-related terms more directly. As a result, even without an
initial velocity gradient, density variation can produce a relatively strong
nonlinear response through induced velocity and temperature fields.

(ii) Linear growth stage ($\rho_L$ > 0.9): $\Delta_{4,2xx-\max}^*=A_2%
\rho_L+B_2$.
The linear branch remains appreciable, with a fitted slope
$A_2=0.090$. This indicates that, after the nonlinear coupling weakens,
density-related terms still provide a strong direct contribution to
$\Delta_{4,2xx\text{-max}}^{*}$.

Figure~\ref{Fig11}(b) shows the temperature-gradient effect on
$\Delta_{4,2xx-\text{max}}^*$, with $T_R=0.2$ fixed. The response changes from
linear to nonlinear as $T_L$ increases.

(i) Linear growth stage (2.1 < $T_L$ < 2.7): $\Delta_{4,2xx-\max}^*=A_1
T_L+B_1$.
Under weak temperature gradients, temperature-related terms and weak coupling
contributions dominate, giving an approximately linear response.

(ii) Exponential growth stage ($T_L$ > 2.7): $\Delta_{4,2xx-\max}^*=A_2
e^{B_2 T_L}+C_2$.
As the temperature gradient increases, the local pressure gradient is
enhanced and induces non-zero velocity gradients. This activates
velocity-gradient-coupled terms in $\Delta_{4,2xx}^{*(2)}$. At the same
time, second-order temperature-gradient terms and temperature--velocity
coupling terms are amplified. The combined effect produces nonlinear
enhancement and leads to exponential growth of
$\Delta_{4,2xx-\text{max}}^*$.

Thus, temperature gradients can drive $\bm{\Delta}_{4,2}^*$ through
second-order coupling mechanisms. Even without an imposed velocity gradient,
temperature gradients can induce high-order TNE responses through
pressure-gradient-mediated velocity variations.

\subsection{Capability to resolve $\bm{\Delta}_2^*$}

\begin{figure}
\centering
\includegraphics[width=0.9\textwidth]{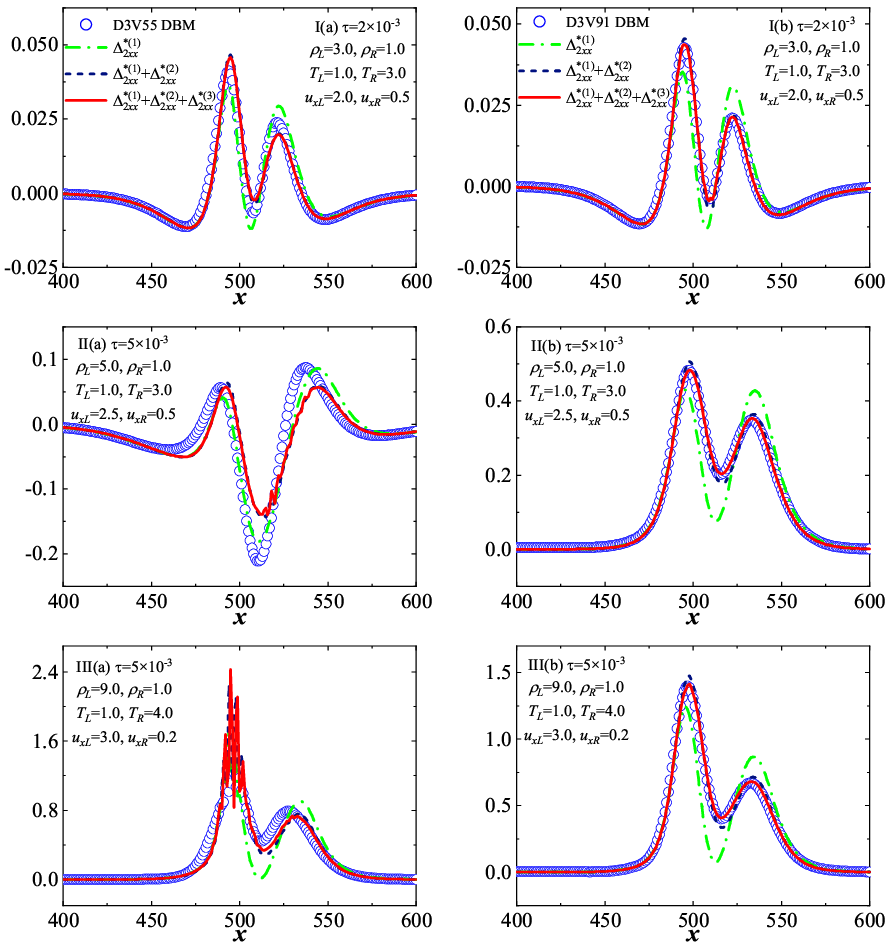}
\caption{Distributions of ${\Delta}_{2xx}^*$ for the weak (I), moderate
(II) and strong (III) cases computed using the D3V55 model (left column)
and the D3V91 model (right column). Blue open circles denote the DBM results. The green dash-dotted, black
dashed and red solid lines denote the first-, second- and third-order
analytical solutions, respectively.
The corresponding times for cases I--III are $t=0.015$, $0.0325$ and $0.03$, respectively.}
\label{Fig12}
\end{figure}

To test the ability of the model to resolve the viscous stress
$\bm{\Delta}_2^*$, we compare the Burnett-level D3V55 model with the
super-Burnett-level D3V91 model under three prescribed cases:

(i) $\rho_L=3, \rho_R=1, T_L=1, T_R=3, u_{xL}=2, u_{xR}=0.5,
L_\rho=L_u=L_T=15, \tau = 2\times10^{-3}$;

(ii) $\rho_L=5, \rho_R=1, T_L=1, T_R=3, u_{xL}=2.5, u_{xR}=0.5,
L_\rho=L_u=L_T=20, \tau = 5\times10^{-3}$;

(iii) $\rho_L=9, \rho_R=1, T_L=1, T_R=4, u_{xL}=3, u_{xR}=0.2,
L_\rho=L_u=L_T=20, \tau = 5\times10^{-3}$.

Figure~\ref{Fig12} compares the distributions of $\Delta_{2xx}^{*}$, and
Table~\ref{Table1} reports the relative contributions of the second- and
third-order components.

\begin{longtable}{C{3cm}C{3cm}C{3cm}C{3cm}}
  \caption{Relative TNE intensities of the viscous stress for the weak (I), moderate (II) and strong (III) nonequilibrium cases.} \label{Table1} \\
  \toprule[1pt]
  \rule{0pt}{4ex}
  & Location & $\dfrac{\Delta_{2xx}^{*(2)}}{\Delta_{2xx}^{*(1)}}$ & $\dfrac{\Delta_{2xx}^{*(3)}}{\Delta_{2xx}^{*(1)}}$ \\
  \addlinespace[1pt]
  \midrule
  \multirow{3}{*}{Case I} & $x=495$ & 0.31 & 0.05 \\
                          & $x=509$ & 0.45 & 0.18 \\
                          & $x=522$ & 0.29 & 0.02 \\
  \midrule
  \multirow{3}{*}{Case II} & $x=499$ & 0.27 & 0.06 \\
                           & $x=515$ & 1.15 & 0.30 \\
                           & $x=534$ & 0.15 & 0.02 \\
  \midrule
  \multirow{3}{*}{Case III} & $x=497$ & 0.21 & 0.06 \\
                            & $x=515$ & 2.38 & 0.68 \\
                            & $x=534$ & 0.18 & 0.04 \\
  \bottomrule[1pt]
\end{longtable}

In case I, the relaxation time and the density, temperature and velocity
gradients are relatively small, corresponding to a weak-TNE regime. The
constitutive expressions for $\Delta_{2xx}^{*}$ show that
$\Delta_{2xx}^{*(1)}$ is primarily driven by the velocity gradient, while
its magnitude is modulated by $\rho$, $T$ and $\tau$. The higher-order terms
include second-order derivatives of macroscopic quantities, such as
$\partial_x^2\rho$, third-order derivatives, such as $\partial_x^3\rho$ and
$\partial_x^3u_x$, and multigradient coupling terms such as
$\partial_xu_x\,\partial_x^2u_x$, $\partial_x\rho\,\partial_x^2u_x$ and
$\partial_xT\,\partial_x^2u_x$. Figures~\ref{Fig12}I(a,b) show that D3V55
deviates from the first-, second- and third-order analytical solutions,
whereas D3V91 is consistent with the third-order analytical solution. At
$x=509$, the relative intensities of the second- and third-order components
reach
$R_\text{TNE}^{(2)/(1)}=\Delta_{2xx}^{*(2)}/\Delta_{2xx}^{*(1)}=0.45$ and
$R_\text{TNE}^{(3)/(1)}=\Delta_{2xx}^{*(3)}/\Delta_{2xx}^{*(1)}=0.18$,
respectively. Thus, even in a weak-TNE regime, a lower-order closure can
fail when the high-order contributions are not negligible.

In case II, the relaxation time, density gradient and velocity gradient are
increased, strengthening nonlinear multigradient coupling. At $x=515$, the
relative intensities increase to
$R_\text{TNE}^{(2)/(1)}=1.15$ and $R_\text{TNE}^{(3)/(1)}=0.30$. The
second-order contribution therefore exceeds the first-order contribution,
while the third-order contribution also becomes important. D3V91 remains
consistent with the third-order analytical solution, whereas D3V55 shows
larger deviations because it lacks the moment relations required to
represent the cross-couplings among gradients at different scales.

In case III, the density, temperature and velocity gradients are further
increased, driving the system into a strongly nonequilibrium regime. At
$x=515$, the relative intensities reach
$R_\text{TNE}^{(2)/(1)}=2.38$ and $R_\text{TNE}^{(3)/(1)}=0.68$, indicating that
high-order components dominate the viscous-stress response. The relative
intensities also vary strongly among different spatial locations, reflecting
the multiscale nature of the TNE response. Figures~\ref{Fig12}III(a,b) show
substantial deviations from the constitutive descriptions for D3V55. The
oscillatory behaviour in the evaluated higher-order profiles is associated
with discontinuities in the macroscopic fields. By contrast, D3V91 resolves
the third-order contribution and remains consistent with the third-order
analytical solution. These results show that resolving $\bm{\Delta}_2^*$
requires the corresponding super-Burnett-level moment constraints when the
second- and third-order contributions become appreciable.

\subsection{Nonequilibrium phase diagram of $\bm{\Delta}_2^*$}

Although $\bm{\Delta}_2^*$ is primarily associated with velocity gradients,
density and temperature gradients also modify its response through explicit
thermodynamic dependences and higher-order coupling terms. We therefore
examine these effects to identify the multigradient-coupling mechanisms
associated with the viscous stress.

\begin{figure}
\centering
\includegraphics[width=1\textwidth]{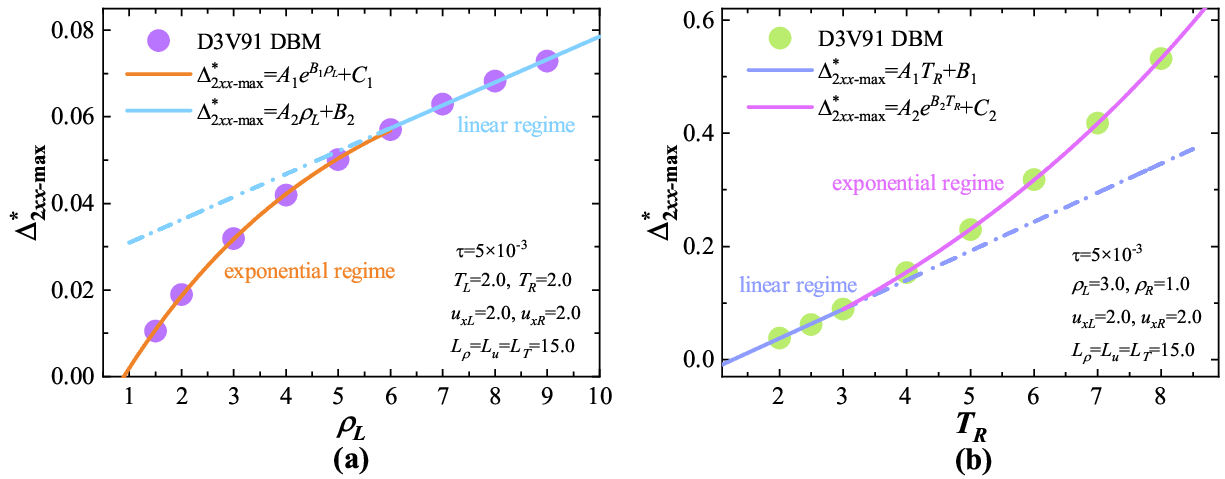}
\caption{Effects of the density gradient (a) and temperature gradient (b) on
the nonequilibrium quantity $\bm{\Delta}_2^*$: linear and exponential
responses, where for (a): $A_1 = 0.095$, $B_1 = 0.153$, $C_1 = -0.089$, $A_2 = 0.005$, $B_2 = 0.026$; for (b): $A_1 = 0.051$, $B_1 = -0.066$, $A_2 = 0.290$, $B_2 = 0.139$, $C_2 = -0.352$.}
\label{Fig13}
\end{figure}

Figure~\ref{Fig13}(a) shows the density effect on
$\Delta_{2xx\text{-max}}^{*}$. As $\rho_L$ increases,
$\Delta_{2xx\text{-max}}^{*}$ exhibits two response regimes. Unlike the
density-gradient effects discussed above, the response of
$\Delta_{2xx}^{*}$ is controlled not only by the relative density variation,
$\bm{\nabla}\rho/\rho$, but also by the explicit density dependence in the
constitutive expressions.

(i) Exponential growth stage (1.5 < $\rho_L$ < 6): $\Delta_{2xx-\max}^*=A_1
e^{B_1\rho_L}+C_1$.
The first-order term $\Delta_{2xx}^{*(1)}$ is proportional to the density,
whereas the higher-order terms contain nonlinear density dependences,
including $\rho^2$, $1/\rho$, $\rho^{-2}$ and $\rho^3$, as well as
density-derivative terms such as $\partial_x^2\rho$ and
$(\partial_x\rho)^2$. In the lower-density regime, the relative density
variation remains sufficiently large to enhance compressibility-related
coupling. Together with the explicit nonlinear density dependences, these
terms amplify the viscous-stress response and produce an approximately
exponential increase in $\Delta_{2xx\text{-max}}^{*}$.

(ii) Linear growth stage ($\rho_L$ > 6): $\Delta_{2xx-\max}^*=A_2\rho_L+B_2$.
The fitted slope of the linear branch is small, $A_2=0.005$, showing that
the density effect on $\bm{\Delta}_2^*$ becomes weak once the nonlinear
density-dependent terms have saturated. The remaining response is governed
mainly by the leading explicit density dependence of the first-order
viscous-stress contribution.

Figure~\ref{Fig13}(b) shows the temperature-gradient effect on
$\Delta_{2xx\text{-max}}^{*}$ by increasing the right-state temperature
$T_R$, with $T_L=1$ fixed. As the temperature gradient increases, the
response changes from an approximately linear dependence to nonlinear
growth.

(i) Linear growth stage (2 < $T_R$ < 3): $\Delta_{2xx-\max}^*=A_1 T_R+B_1$.
For weak temperature gradients, the higher-order contributions remain small.
The response is therefore dominated by the first-order term, and the induced
velocity-gradient response is weak.

(ii) Exponential nonlinear growth stage ($T_R$ > 3): $\Delta_{2xx-%
\max}^*=A_2 e^{B_2 T_R}+C_2$.
As the temperature gradient increases, multigradient coupling becomes more
pronounced. The second-order contribution contains terms involving $T^2$,
$T$, $(\partial_xT)^2$ and mixed gradients. The third-order contribution
further introduces couplings among temperature, velocity and density
gradients, together with curvature-related terms such as
$\partial_xT\,\partial_x^2u_x$,
$\partial_xT\,\partial_x^2\rho$ and
$\partial_xu_x(\partial_xT)^2$, with an overall scaling proportional to
$\tau^3$. These coupled terms amplify the viscous-stress response and lead
to an approximately exponential increase in
$\Delta_{2xx\text{-max}}^{*}$.

Thus, density and temperature gradients modify $\bm{\Delta}_2^*$ through
different mechanisms. The density effect combines explicit density
dependence with relative-gradient-related corrections, whereas the
temperature effect mainly enters through mixed-gradient and curvature
couplings. These results show that the viscous stress is not determined by
the velocity gradient alone once higher-order TNE contributions become
appreciable.

\subsection{Capability to resolve $\bm{\Delta}_{3,1}^*$}

To test the capability of the model to resolve heat flux $\bm{\Delta}_{3,1}^*$, we consider
three cases with different TNE intensities:

(i) $\rho_L=1, \rho_R=3, T_L=2, T_R=2, u_{xL}=0, u_{xR}=0,
L_\rho=L_u=L_T=20, \tau = 3 \times 10^{-3}$;

(ii) $\rho_L=3, \rho_R=1, T_L=4, T_R=4, u_{xL}=0.1, u_{xR}=0.1,
L_\rho=L_u=L_T=20, \tau = 5 \times 10^{-3}$;

(iii) $\rho_L=4, \rho_R=4, T_L=4, T_R=1, u_{xL}=0.1, u_{xR}=0.1,
L_\rho=L_u=L_T=20, \tau = 5 \times 10^{-3}$.

\begin{figure}
\centering
\includegraphics[width=0.9\textwidth]{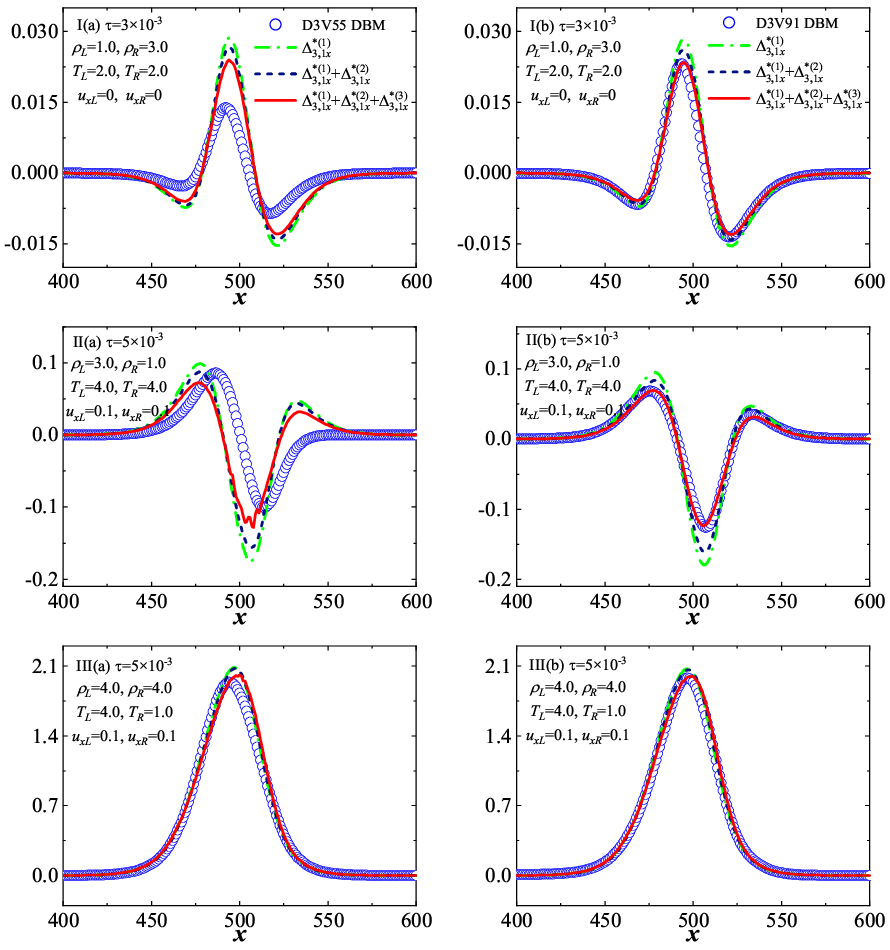}
\caption{Distributions of ${\Delta}_{3,1x}^*$ for the weak (I), moderate
(II), and strong (III) cases computed using the D3V55 model (left column)
and the D3V91 model (right column). Blue open circles denote the DBM results. The green dash-dotted, black
dashed and red solid lines denote the first-, second- and third-order
analytical solutions, respectively.
 The corresponding times for cases I--III are $t=0.015$, $0.018$, and $0.016$,
respectively.}
\label{Fig14}
\end{figure}

Figure~\ref{Fig14} compares $\Delta_{3,1x}^{*}$ obtained with the
Burnett-level D3V55 model and the super-Burnett-level D3V91 model.
In case I, no temperature gradient is imposed initially, so the early
heat-flux response is initiated by the density gradient. The density
gradient first drives the second-order contribution
$\Delta_{3,1x}^{*(2)}$ and induces a velocity-gradient response. During the
subsequent evolution, this coupled response generates a temperature gradient,
so the first-order contribution $\Delta_{3,1x}^{*(1)}$ becomes dominant at
the analysed time. The maximum relative intensities of the second- and
third-order components are
$R_\text{TNE}^{(2)/(1)}=0.08$ at $x=495$ and
$R_\text{TNE}^{(3)/(1)}=0.12$ at $x=469$, respectively. Thus, although the
instantaneous heat flux is dominated by the first-order contribution, its
formation is affected by the density-gradient-induced higher-order pathway.

Even in this weakly nonequilibrium regime, D3V55 shows visible deviations
because it lacks the moment constraints required to represent the
higher-order derivative and coupling terms involved in this pathway. By
contrast, D3V91 is consistent with the third-order truncated analytical
expression. This comparison indicates that density-gradient-induced
heat-flux responses can be sensitive to high-order moment closure, even when
the first-order contribution dominates the total signal.

\begin{longtable}{C{3cm}C{3cm}C{3cm}C{3cm}}
  \caption{Relative nonequilibrium intensities of the heat flux for weak (I), moderate (II), and strong (III) nonequilibrium conditions.} \label{Table2} \\
  \toprule[1pt]  
  \rule{0pt}{4ex}
  & corresponding location & $\dfrac{\Delta_{3,1x}^{*(2)}}{\Delta_{3,1x}^{*(1)}}$ & $\dfrac{\Delta_{3,1x}^{*(3)}}{\Delta_{3,1x}^{*(1)}}$ \\
  \addlinespace[1pt]
  \midrule
  \multirow{3}{*}{Case I} & $x=469$ & 0.07 & 0.12 \\
                          & $x=495$ & 0.08 & 0.1 \\
                          & $x=522$ & 0.08 & 0.07 \\
  \midrule
  \multirow{3}{*}{Case II} & $x=478$ & 0.12 & 0.15 \\
                           & $x=506$ & 0.1  & 0.21 \\
                           & $x=533$ & 0.08 & 0.25 \\
  \midrule
  Case III & $x=497$ & 0.005 & 0.03 \\
  \toprule[1pt]
\end{longtable}

In case II, the relaxation time and macroscopic amplitudes are increased,
which strengthens multiscale coupling among gradients and transport orders.
The maximum relative intensities increase to
$R_\text{TNE}^{(2)/(1)}=0.12$ at $x=478$ and
$R_\text{TNE}^{(3)/(1)}=0.25$ at $x=533$. The third-order contribution thus
becomes more important, although the first-order heat flux remains dominant.
The D3V55 model cannot represent the enhanced multigradient coupling and
therefore deviates from the constitutive descriptions
(Figs.~\ref{Fig14}II(a,b)). By contrast, D3V91 remains consistent with the
third-order analytical solution, showing its advantage for coupled heat-flux
transport.

In case III, the density gradient is removed while the temperature gradient
is increased. Although the absolute TNE intensity is about 20 times that in
case II, the relative higher-order contributions are reduced:
$R_\text{TNE}^{(2)/(1)}=0.005$ and $R_\text{TNE}^{(3)/(1)}=0.03$ at $x=497$.
The system therefore returns to a first-order heat-flux regime. Higher-order
coupling is weak because heat transport broadens the macroscopic interface
and reduces the effective TNE driving strength. In this regime, the
discrepancy of D3V55 decreases, while D3V91 still represents the third-order
heat-flux behaviour consistently. These results show that resolving
$\bm{\Delta}_{3,1}^*$ requires super-Burnett-level moment constraints when
density-gradient-induced or multigradient coupling contributions become
appreciable.

\subsection{Nonequilibrium phase diagram of $\bm{\Delta}_{3,1}^*$}

\begin{figure}
\centering
\includegraphics[width=1\textwidth]{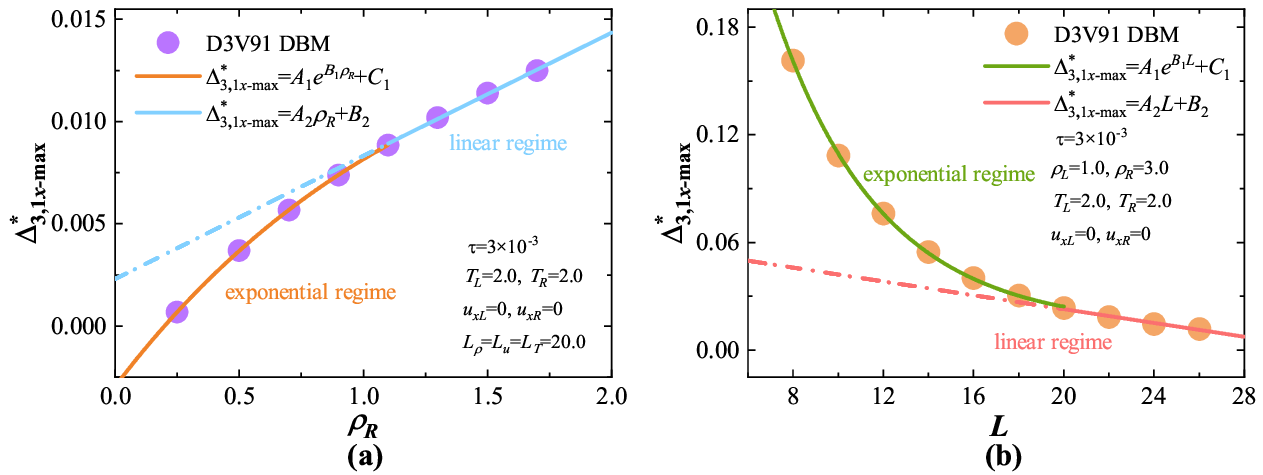}
\caption{Effects of the density gradient (a) and transition-layer width (b) on
the heat flux $\bm \Delta_{3,1}^{*}$: exponential and linear responses, where for (a): $A_1 = 0.013$, $B_1 = 0.917$, $C_1 = -0.005$, $A_2 = 0.006$, $B_2 = 0.002$; for (b): $A_1 = -0.703$, $B_1 = -0.110$, $C_1 = 0.012$, $A_2 = -0.002$, $B_2 = 0.061$.}
\label{Fig15}
\end{figure}

Figure~\ref{Fig15}(a) shows the density-gradient effect on
$\Delta_{3,1x\text{-max}}^{*}$ when no temperature or velocity gradient is
imposed initially and the left-state density is fixed at $\rho_L=0.2$. Among
the four TNE quantities considered here, this is the weakest density-induced
response: the fitted values remain at the order of $10^{-2}$.

(i) Exponential growth stage (0.25 < $\rho_R$ < 1.1): $\Delta_{3,1x-%
\max}^*=A_1 e^{B_1\rho_R}+C_1$.
The exponential branch is mild. Because $\bm{\Delta}_{3,1}^{*}$ is aheat-flux-related vector moment, a pure density gradient affects it mainlythrough an indirect pathway: density variation first induces higher-ordernonequilibrium and regular velocity or temperature responses, which then feedback into the heat-flux component. This indirect pathway explains thenonlinear form of the fit.

(ii) Linear growth stage ($\rho_R$ > 1.1): $\Delta_{3,1x-\max}^*=A_2\rho_R+B_2$.
The fitted slope is only $A_2=0.006$, confirming that the density-gradient
effect remains weak in the high-density branch. Thus, for
$\bm{\Delta}_{3,1}^{*}$, density variation acts as a secondary trigger
rather than a dominant driving mechanism.

We next isolate the transition-layer-width effect of the imposed density
interface by varying $L_\rho=L\Delta x$, where $L$ denotes the dimensionless
interface width in grid units. In this test, $T_L=T_R$ and
$u_{xL}=u_{xR}=0$, so temperature and velocity gradients are not imposed
initially but may be generated during the subsequent evolution.
Figure~\ref{Fig15}(b) shows the variation of
$\Delta_{3,1x\text{-max}}^{*}$ with $L$ for
$\tau=3\times10^{-3}$. Increasing $L$ broadens the density interface and
reduces the imposed density gradient. Consequently, the heat-flux response
decreases monotonically and exhibits two response regimes.

(i) Exponential decay stage (8 < $L$ < 20): $\Delta_{3,1x-\max}^*=A_1 e^{B_1
L}+C_1$.
In narrow density transition layers, the imposed density gradient is strong.
It induces localized velocity and temperature responses during the early
evolution, so the resulting gradients remain concentrated near the interface
and couple effectively. Increasing $L$ rapidly weakens the imposed density
gradient and reduces the induced multigradient coupling. This produces an
approximately exponential decay of
$\Delta_{3,1x\text{-max}}^{*}$.

(ii) Linear decay stage ($L$ > 20): $\Delta_{3,1x-\max}^*=A_2 L+B_2$.
In wide transition layers, the imposed density gradient is weaker, and the
induced velocity and temperature responses are more diffuse. The nonlinear
coupling is then reduced, and the response is governed mainly by the gradual
decrease of the density-gradient strength with interface width. As a result,
$\Delta_{3,1x\text{-max}}^{*}$ decays approximately linearly with $L$.

A similar decrease with increasing transition-layer width is observed for
$\bm{\Delta}_{3}^{*}$, $\bm{\Delta}_{4,2}^{*}$ and $\bm{\Delta}_{2}^{*}$.
This common behaviour indicates that the transition-layer width controls
high-order TNE mainly by regulating the strength, localization and overlap of
the imposed and induced macroscopic gradients. Narrow transition layers
enhance nonlinear multigradient coupling, whereas wide transition layers
weaken this coupling and lead to a more nearly linear response.

\section{Conclusions and outlook}
\label{Conclusions}

This study develops a three-dimensional super-Burnett-level discrete
Boltzmann model for high-order thermodynamic nonequilibrium (TNE) in
compressible flows. The central objective is to connect kinetic moment
closure with high-order non-conserved moments and their
multigradient-coupled responses.

First, we derived the kinetic moment relations required for high-order TNE
from the Chapman--Enskog expansion. These relations specify the moment
constraints needed to describe third-order contributions to
$\bm{\Delta}_2^*$ and $\bm{\Delta}_{3,1}^*$, together with second-order
contributions to $\bm{\Delta}_3^*$ and $\bm{\Delta}_{4,2}^*$. The resulting
moment set is
$\boldsymbol{\Phi} = (\mathbf{M}_0, \mathbf{M}_1, \mathbf{M}_{2,0}, \mathbf{M}_2,
\mathbf{M}_{3,1}, \mathbf{M}_3, \mathbf{M}_{4,2}, \mathbf{M}_4,
\mathbf{M}_{5,3}, \mathbf{M}_5, \mathbf{M}_{6,4})$.
Based on this set, we constructed two D3V91 discrete velocity realizations
and obtained the corresponding three-dimensional nonlinear constitutive
relations. Compared with two-dimensional formulations, the
three-dimensional construction contains more independent moment components
and involves richer coupling among gradients, curvature terms and
high-order fluxes.

Second, one-dimensional compressible benchmark problems were used to assess
the macroscopic consistency of the model. The D3V91 model reproduces the
main Riemann wave structures, including shocks, rarefaction waves and
contact discontinuities. These tests assess the implementation of the
discrete velocity model in one-dimensional strong-gradient flows, while not
constituting a complete validation for fully three-dimensional flow
configurations.

Third, we used the model to examine high-order TNE quantities. The results
show that the Burnett-level D3V55 model can describe some lower-order TNE
responses, but becomes insufficient when high-order contributions become
comparable to, or larger than, the leading-order terms. By contrast, the
D3V91 model satisfies the additional moment constraints required to
represent third-order contributions to $\bm{\Delta}_2^*$ and
$\bm{\Delta}_{3,1}^*$, and second-order contributions to
$\bm{\Delta}_3^*$ and $\bm{\Delta}_{4,2}^*$. These comparisons show that
the ability to resolve a target TNE quantity depends not only on the number
of discrete velocities, but more fundamentally on whether the required
kinetic moment relations are satisfied.

Fourth, the parametric response analysis shows that high-order TNE
quantities are not driven by a single macroscopic gradient. Instead, they
arise from coupled leading gradients, secondary gradients, gradient
products, curvature terms and transition-layer widths. Depending on the
relative strength of these driving factors, the TNE response may change from
an approximately linear dependence to nonlinear or exponential growth, and
may return to an approximately linear regime. The density-gradient analysis
further shows that the relevant driving measure is often the relative
density variation, $\bm{\nabla}\rho/\rho$, rather than the absolute density
gradient alone. These results identify multigradient coupling as a key
mechanism controlling high-order TNE in strong-gradient compressible flows.

The present model has several limitations. The single-relaxation-time
formulation assumes a common relaxation rate for all kinetic modes, whereas
realistic nonequilibrium flows may involve mode-dependent relaxation
associated with momentum transport, heat transport, internal-energy
exchange, chemical reactions and multicomponent diffusion. The numerical
tests are also restricted mainly to one-dimensional configurations, although
the discrete velocity model and the moment relations are three-dimensional.
Further assessment is therefore required for genuinely three-dimensional
flows with shock interactions, vortical structures and strong thermal or
compositional gradients. In addition, the stability properties of the
present high-order velocity set and its coupling with numerical schemes
remain to be analysed systematically.

Future work should proceed in three directions. First, moment closures
beyond the present super-Burnett level should be developed for more extreme
nonequilibrium regimes. Second, multiple-relaxation-time and mode-dependent
collision models should be constructed to represent the distinct relaxation
rates of different kinetic processes. Third, systematic stability analyses,
including von Neumann analysis, should be carried out to identify the
constraints imposed by the discrete velocity set, relaxation parameters and
numerical schemes. These developments will extend high-order DBMs to more
complex compressible, rarefied, reacting and multicomponent flows.

\linespread{1.0}

\section*{Acknowledgements}

This work was supported by the National Natural Science Foundation of China
(Grant Nos. U2242214, 11875001 and 12172061), the Hebei Outstanding Youth Science
Foundation (Grant No. A2023409003), the Natural Science Foundation of Fujian
Province (Grant No. 2026J001415), and the Fujian Provincial Special Funds
for Education and Research in Provincial Units (Grant No. K3-949).

\appendix

\section{Discrete velocity sets and high-order constitutive terms}

This appendix lists the discrete velocity sets and the analytical
constitutive terms used in the D3V91 models. Tables~\ref{TableI} and
\ref{TableII} give the two 91-velocity realizations used in the numerical
tests. Tables~\ref{TableIII}--\ref{TableVII} provide the first-, second-
and third-order TNE expressions used to compare the DBM results with the
Chapman--Enskog constitutive relations.

\begin{center}
\begin{longtable}{|C{1.3cm}|C{2.9cm}|C{1.3cm}|C{2.9cm}|C{1.3cm}|C{2.9cm}|}
\caption{Discrete velocity set D3V91(I), with 91 discrete velocities.} \label{TableI}\\
\hline
\textbf{Index} & \textbf{$(v_x, v_y, v_z)$} & \textbf{Index} & \textbf{$(v_x, v_y, v_z)$} & \textbf{Index} & \textbf{$(v_x, v_y, v_z)$} \\
\hline
\endfirsthead

\hline
\textbf{Index} & \textbf{$(v_x, v_y, v_z)$} & \textbf{Index} & \textbf{$(v_x, v_y, v_z)$} & \textbf{Index} & \textbf{$(v_x, v_y, v_z)$} \\
\hline
\endhead

\endlastfoot

1   & (0,0,0)             & 32  & ($-$1,2,0)        & 62  & (0,$-$2,$-$2) \\ \hline
2   & (1,0,0)             & 33  & (1,$-$2,0)        & 63  & (0,2,$-$2)  \\ \hline
3   & ($-$1,0,0)          & 34  & ($-$2,1,0)        & 64  & (2,1,3)   \\ \hline
4   & (0,1,0)             & 35  & (2,$-$1,0)        & 65  & ($-$2,$-$1,$-$3)\\ \hline
5   & (0,$-$1,0)          & 36  & (0,2,1)           & 66  & (3,1,2)   \\ \hline
6   & (0,0,1)             & 37  & (0,$-$2,$-$1)     & 67  & ($-$3,$-$1,$-$2)\\ \hline
7   & (0,0,$-$1)          & 38  & (0,1,2)           & 68  & (3,1,$-$2)  \\ \hline
8   & (1,1,0)             & 39  & (0,$-$1,$-$2)     & 69  & ($-$3,$-$1,2) \\ \hline
9   & ($-$1,1,0)          & 40  & (0,$-$1,2)        & 70  & (2,1,$-$3)  \\ \hline
10  & ($-$1,$-$1,0)         & 41  & (0,1,$-$2)        & 71  & ($-$2,$-$1,3) \\ \hline
11  & (1,$-$1,0)          & 42  & (0,$-$2,1)        & 72  & (3,2,1)   \\ \hline
12  & (1,0,1)             & 43  & (0,2,$-$1)        & 73  & ($-$3,$-$2,$-$1)\\ \hline
13  & (1,0,$-$1)          & 44  & (2,0,1)           & 74  & (2,3,1)   \\ \hline
14  & ($-$1,0,$-$1)       & 45  & ($-$2,0,$-$1)     & 75  & ($-$2,$-$3,$-$1)\\ \hline
15  & ($-$1,0,1)          & 46  & (1,0,2)           & 76  & ($-$2,3,1)  \\ \hline
16  & (0,1,1)             & 47  & ($-$1,0,$-$2)     & 77  & (2,$-$3,$-$1) \\ \hline
17  & (0,$-$1,1)          & 48  & ($-$1,0,2)        & 78  & ($-$3,2,1)  \\ \hline
18  & (0,$-$1,$-$1)       & 49  & (1,0,$-$2)        & 79  & (3,$-$2,$-$1) \\ \hline
19  & (0,1,$-$1)          & 50  & ($-$2,0,1)        & 80  & (1,2,3)   \\ \hline
20  & (1,1,1)             & 51  & (2,0,$-$1)        & 81  & ($-$1,$-$2,$-$3)\\ \hline
21  & ($-$1,$-$1,$-$1)    & 52  & (2,2,0)           & 82  & (1,3,2)   \\ \hline
22  & ($-$1,1,1)          & 53  & ($-$2,2,0)        & 83  & ($-$1,$-$3,$-$2)\\ \hline
23  & (1,$-$1,$-$1)       & 54  & ($-$2,$-$2,0)     & 84  & (1,3,$-$2)  \\ \hline
24  & ($-$1,$-$1,1)       & 55  & (2,$-$2,0)        & 85  & ($-$1,$-$3,2) \\ \hline
25  & (1,1,$-$1)          & 56  & (2,0,2)           & 86  & (1,2,$-$3)  \\ \hline
26  & (1,$-$1,1)          & 57  & (2,0,$-$2)        & 87  & ($-$1,$-$2,3) \\ \hline
27  & ($-$1,1,$-$1)       & 58  & ($-$2,0,$-$2)     & 88  & ($-$2,1,$-$1) \\ \hline
28  & (2,1,0)             & 59  & ($-$2,0,2)        & 89  & ($-$2,$-$2,2) \\ \hline
29  & ($-$2,$-$1,0)      & 60  & (0,2,2)           & 90  & (2,$-$3,$-$3) \\ \hline
30  & (1,2,0)             & 61  & (0,$-$2,2)        & 91  & (2,4,4)   \\ \hline
31  & ($-$1,$-$2,0)       &     &           &     &           \\ \hline
\end{longtable}

\begin{longtable}{|C{1.3cm}|C{2.9cm}|C{1.3cm}|C{2.9cm}|C{1.3cm}|C{2.9cm}|}
\caption{Discrete velocity set D3V91(II), with 91 discrete velocities.} \label{TableII} \\
\hline
\textbf{Index} & \textbf{$(v_x, v_y, v_z)$} & \textbf{Index} & \textbf{$(v_x, v_y, v_z)$} & \textbf{Index} & \textbf{$(v_x, v_y, v_z)$} \\
\hline
\endfirsthead

\hline
\textbf{Index} & \textbf{$(v_x, v_y, v_z)$} & \textbf{Index} & \textbf{$(v_x, v_y, v_z)$} & \textbf{Index} & \textbf{$(v_x, v_y, v_z)$} \\
\hline
\endhead

\endlastfoot

1   & (0, 0, 0)                 & 32  & (0, $-$2, 0)                & 62  & ($-$3.1, 3.5, 3.1)          \\ \hline
2   & (0.7, 0, 0)               & 33  & (0, 0, 2)                  & 63  & ($-$3.1, $-$3.5, $-$3.1)        \\ \hline
3   & ($-$0.7, 0, 0)              & 34  & (0, 0, $-$2)                 & 64  & ($-$3.1, 3.5, $-$3.1)         \\ \hline
4   & (0, 0.7, 0)               & 35  & (1.1, 3, 0)                & 65  & ($-$3.1, $-$3.5, 3.1)         \\ \hline
5   & (0, $-$0.7, 0)              & 36  & ($-$1.1, 3, 0)               & 66  & (3.5, 5, 5)               \\ \hline
6   & (0, 0, 0.7)               & 37  & (1.1, $-$3, 0)               & 67  & (3.5, $-$4.6, 4.6)          \\ \hline
7   & (0, 0, $-$0.7)              & 38  & ($-$1.1, $-$3, 0)              & 68  & (3.5, 4.2, $-$4.2)          \\ \hline
8   & (0.7, 0.7, 0)             & 39  & (0, 2.1, 2.8)              & 69  & (3.5, $-$3.8, $-$3.8)         \\ \hline
9   & ($-$0.7, 0.7, 0)            & 40  & (0, $-$2.1, 2.8)             & 70  & ($-$3.5, 3.8, 3.8)          \\ \hline
10  & ($-$0.7, $-$0.7, 0)           & 41  & (0, 2.1, $-$2.8)             & 71  & ($-$3.5, $-$4.2, 4.2)         \\ \hline
11  & (0.7, $-$0.7, 0)            & 42  & (0, $-$2.1, $-$2.8)            & 72  & ($-$3.5, 4.6, $-$4.6)         \\ \hline
12  & (0.7, 0, 0.7)             & 43  & (3, 0, 3.5)                     & 73  & ($-$3.5, $-$5, $-$5)            \\ \hline
13  & (0.7, 0, $-$0.7)            & 44  & ($-$3, 0, 3.5)               & 74  & (7.2, 0, 0)               \\ \hline
14  & ($-$0.7, 0, $-$0.7)           & 45  & (3, 0, $-$3.5)             & 75  & ($-$7.2, 0, 0)              \\ \hline
15  & ($-$0.7, 0, 0.7)            & 46  & ($-$3, 0, $-$3.5)            & 76  & (0, 7.2, 0)               \\ \hline
16  & (0, 0.7, 0.7)             & 47  & (5, 3.1, 1.7)                  & 77  & (0, $-$7.2, 0)              \\ \hline
17  & (0, $-$0.7, 0.7)            & 48  & ($-$4.6, 3.1, 2)             & 78  & (0, 0, 7.2)               \\ \hline
18  & (0, $-$0.7, $-$0.7)           & 49  & (4.2, $-$3.1, 2.3)         & 79  & (0, 0, $-$7.2)              \\ \hline
19  & (0, 0.7, $-$0.7)            & 50  & ($-$3.8, $-$3.1, 2.6)        & 80  & (5.5, 5, 6)               \\ \hline
20  & (1.9, 1.9, 1.9)           & 51  & (3.8, 3.1, $-$2.6)             & 81  & (5.5, 5, $-$6)              \\ \hline
21  & ($-$1.9, $-$1.9, $-$1.9)        & 52  & ($-$4.2, $-$3.1, $-$2.3)         & 82  & ($-$5.5, $-$5, 6)             \\ \hline
22  & ($-$1.9, 1.9, 1.9)          & 53  & (4.6, 3.1, $-$2)             & 83  & ($-$5.5, $-$5, $-$6)            \\ \hline
23  & (1.9, $-$1.9, $-$1.9)         & 54  & ($-$5, $-$3.1, $-$1.7)           & 84  & (6.5, 6.5, 0)             \\ \hline
24  & ($-$1.9, $-$1.9, 1.9)         & 55  & (3.1, 3.5, 3.1)            & 85  & (6.5, $-$6.5, 0)            \\ \hline
25  & (1.9, 1.9, $-$1.9)          & 56  & (3.1, $-$3.5, 3.1)           & 86  & ($-$6.5, 6.5, 0)            \\ \hline
26  & (1.9, $-$1.9, 1.9)          & 57  & ($-$3.1, 3.5, 3.1)           & 87  & ($-$6.5, $-$6.5, 0)           \\ \hline
27  & ($-$1.9, 1.9, $-$1.9)         & 58  & (3.1, 3.5, $-$3.1)         & 88  & ($-$4, 7, $-$5)               \\ \hline
28  & (2, 0, 0)                 & 59  & (3.1, $-$3.5, $-$3.1)          & 89  & ($-$4, 9, 6)                \\ \hline
29  & ($-$2, 0, 0)                & 60  & ($-$3.1, $-$3.5, $-$3.1)     & 90  & (2, $-$7, $-$8)               \\ \hline
30  & (0, 2, 0)                 & 61  & ($-$3.1, 3.5, $-$3.1)          & 91  & (6, $-$9, 7)                \\ \hline
31  & (0, $-$2, 0)                &     &                             &     &                           \\ \hline
\end{longtable}
\end{center}

The expressions for the first-order $\bm{\Delta}_{3}^{*(1)}$ and $\bm{\Delta}_{4,2}^{*(1)}$ in three-dimensions are presented in Table \ref{TableIII}.

\begin{longtable}{|c|>{\centering\arraybackslash}m{12cm}|}
\caption{Expressions for  $\bm{\Delta}_{3}^{*(1)}$ and $\bm{\Delta}_{4,2}^{*(1)}$.} \label{TableIII}\\ \hline
\textbf{TNE} & \textbf{Formula} \\ \hline
\endfirsthead

\hline
\textbf{TNE} & \textbf{Formula} \\ \hline
\endhead

\endlastfoot
\( \Delta_{3xxx}^{*(1)} \) &
\(-3 \tau \rho R^2 T \partial_x T\) \\ \hline
\( \Delta_{3xxy}^{*(1)} \) &
\(-\tau \rho R^2 T \partial_y T\) \\ \hline
\( \Delta_{3xxz}^{*(1)} \) &
\(-\tau \rho R^2 T \partial_z T\) \\ \hline
\( \Delta_{3xyy}^{*(1)} \) &
\(-\tau \rho R^2 T \partial_x T\) \\ \hline
\( \Delta_{3xyz}^{*(1)} \) &
\(0\) \\ \hline
\( \Delta_{3xzz}^{*(1)} \) &
\(-\tau \rho R^2 T \partial_x T\) \\ \hline
\( \Delta_{3yyy}^{*(1)} \) &
\(-3 \tau \rho R^2 T \partial_y T\) \\ \hline
\( \Delta_{3yyz}^{*(1)} \) &
\(-\tau \rho R^2 T \partial_z T\) \\ \hline
\( \Delta_{3yzz}^{*(1)} \) &
\(-\tau \rho R^2 T \partial_y T\) \\ \hline
\( \Delta_{3zzz}^{*(1)} \) &
\(-3 \tau \rho R^2 T \partial_z T\) \\ \hline

\( \Delta_{4,2xx}^{*(1)} \) &
\( -n_3^{-1} n_7 \tau \rho R^2 T^2\left(n_2 \partial_x u_x-\partial_y u_y-\partial_z u_z\right)\) \\ \hline
\( \Delta_{4,2xy}^{*(1)} \) &
\( -\frac{1}{2} n_7 \tau \rho R^2 T^2\left(\partial_y u_x+\partial_x u_y\right) \) \\ \hline
\( \Delta_{4,2xz}^{*(1)} \) &
\( -\frac{1}{2} n_7 \tau \rho R^2 T^2\left(\partial_z u_x+\partial_x u_z\right)\) \\ \hline
\( \Delta_{4,2yy}^{*(1)} \) &
\( -n_3^{-1} n_7 \tau \rho R^2 T^2\left(-\partial_x u_x+n_2 \partial_y u_y-\partial_z u_z\right)\) \\ \hline
\( \Delta_{4,2yz}^{*(1)} \) &
\( -\frac{1}{2} n_7 \tau \rho R^2 T^2\left(\partial_z u_y+\partial_y u_z\right)\) \\ \hline
\( \Delta_{4,2zz}^{*(1)} \) &
\( -n_3^{-1} n_7 \tau \rho R^2 T^2\left(-\partial_x u_x-\partial_y u_y+n_2 \partial_z u_z\right)\) \\ \hline
$n_a$ & $n+a$ \\ \hline
\end{longtable}

The expressions for the second-order $\bm{\Delta}_{3}^{*(2)}$ and $\bm{\Delta}_{4,2}^{*(2)}$ in three-dimensions are presented in Table \ref{TableIV}, with the associated terms $S_{1}, \dots, S_{68}$ listed in Table
\ref{TableV}.

\begin{longtable}{|c|>{\centering\arraybackslash}m{12cm}|}
\caption{Expressions for $\bm{\Delta}_{3}^{*(2)}$ and $\bm{\Delta}_{4,2}^{*(2)}$.} \label{TableIV}\\ \hline
\textbf{TNE} & \textbf{Formula} \\ \hline
\endfirsthead

\hline
\textbf{TNE} & \textbf{Formula} \\ \hline
\endhead

\endlastfoot
\( \Delta_{3xxx}^{*(2)} \) &
\( 6 n_3^{-1} \tau^2 \rho R^2 T\left(T S_{1}+\partial_x T S_{2}+n_3 S_{3}\right) \) \\ \hline
\( \Delta_{3xxy}^{*(2)} \) &
\( 2 n_3^{-1} \tau^2 \rho R^2 T\left(T S_{4}+\partial_y T S_{5}+n_3 S_{6}\right)\) \\ \hline
\( \Delta_{3xxz}^{*(2)} \) &
\( 2 n_3^{-1} \tau^2 \rho R^2 T\left(T S_{7}+\partial_z T S_{8}+n_3 S_{9}\right)\) \\ \hline
\( \Delta_{3xyy}^{*(2)} \) &
\( 2 n_3^{-1} \tau^2 \rho R^2 T\left(T S_{10}+\partial_x T S_{11}+n_3 S_{12}\right)\) \\ \hline
\( \Delta_{3xyz}^{*(2)} \) &
\( 2 \tau^2 \rho R^2 T\left(T S_{13}+S_{14}\right)\) \\ \hline
\( \Delta_{3xzz}^{*(2)} \) &
\( 2 n_3^{-1} \tau^2 \rho R^2 T\left(T S_{15}+\partial_x T S_{16}+n_3 S_{17}\right)\) \\ \hline
\( \Delta_{3yyy}^{*(2)} \) &
\( 6 n_3^{-1} \tau^2 \rho R^2 T\left(T S_{18}+\partial_y T S_{19}+n_3 S_{20}\right)\) \\ \hline
\( \Delta_{3yyz}^{*(2)} \) &
\( 2 n_3^{-1} \tau^2 \rho R^2 T\left(T S_{21}+\partial_z T S_{22}+n_3 S_{23}\right)\) \\ \hline
\( \Delta_{3yzz}^{*(2)} \) &
\( 2 n_3^{-1} \tau^2 \rho R^2 T\left(T S_{24}+\partial_y T S_{25}+n_3 S_{26}\right)\) \\ \hline
\( \Delta_{3zzz}^{*(2)} \) &
\( 6 n_3^{-1} \tau^2 \rho R^2 T\left(T S_{27}+\partial_z T S_{28}+n_3 S_{29}\right)\) \\ \hline

\( \Delta_{4,2xx}^{*(2)} \) &
\( -n_3^{-2} \tau^2 \rho^{-1} R^2 T[n_3 n_7 \rho R T^2 S_{30}-n_3^2 \rho^2 R T S_{31}+\rho^2 T(S_{32}+S_{33}-n_3 S_{34}-n_3 S_{35})-n_3 n_7 \rho^2 R S_{36}-n_3 n_7 R T^2 S_{37}]\) \\ \hline
\( \Delta_{4,2xy}^{*(2)} \) &
\( -n_3^{-1} \tau^2 \rho^{-1} R^2 T(n_3 n_7 R T^2 S_{38}-n_3 n_7 \rho^2 R S_{39}-\rho^2 T \partial_y u_x S_{40}-\rho^2 T \partial_x u_y S_{41}-n_3 \rho^2 T S_{42})\) \\ \hline
\( \Delta_{4,2xz}^{*(2)} \) &
\( -n_3^{-1} \tau^2 \rho^{-1} R^2 T(n_3 n_7 R T^2 S_{43}-n_3 n_7 \rho^2 R S_{44}-\rho^2 T \partial_z u_x S_{45}-\rho^2 T \partial_x u_z S_{46}-n_3 \rho^2 T S_{47})\) \\ \hline
\( \Delta_{4,2yy}^{*(2)} \) &
\( -n_3^{-2} \tau^2 \rho^{-1} R^2 T[n_3 n_7 \rho R T^2 S_{48}-n_3^2 \rho^2 R T S_{49}+\rho^2 T(S_{50}+S_{51}-n_3 S_{52}-n_3 S_{53})-n_3 n_7 \rho^2 R S_{54}-n_3 n_7 R T^2 S_{55}]\) \\ \hline
\( \Delta_{4,2yz}^{*(2)} \) &
\( -n_3^{-1} \tau^2 \rho^{-1} R^2 T(n_3 n_7 R T^2 S_{56}-n_3 n_7 \rho^2 R S_{57}-\rho^2 T \partial_z u_y S_{58}-\rho^2 T \partial_y u_z S_{59}-n_3 \rho^2 T S_{60})\) \\ \hline
\( \Delta_{4,2zz}^{*(2)} \) &
\( -n_3^{-2} \tau^2 \rho^{-1} R^2 T[n_3 n_7 \rho R T^2 S_{61}-n_3^2 \rho^2 R T S_{62}+\rho^2 T(S_{63}+S_{64}-n_3 S_{65}-n_3 S_{66})-n_3 n_7 \rho^2 R S_{67}-n_3 n_7 R T^2 S_{68}]\) \\ \hline
\end{longtable}

\begin{longtable}{|c|>{\centering\arraybackslash}m{12.2cm}|}
\caption{Expressions for $S_i$, $i=1,...,68$.} \label{TableV}\\ \hline
\textbf{TNE} & \textbf{Formula} \\ \hline
\endfirsthead

\hline
\textbf{TNE} & \textbf{Formula} \\ \hline
\endhead

\endlastfoot
\( S_{1} \) & \( n_1 \partial^2_x u_x-2 \partial^2_{xy} u_y-2 \partial^2_{xz} u_z \) \\ \hline
\( S_{2} \) & \( (5+3 n) \partial_x u_x-4 \partial_y u_y-4 \partial_z u_z \) \\ \hline
\( S_{3} \) & \( \partial_y T \partial_y u_x+\partial_z T \partial_z u_x \) \\ \hline
\( S_{4} \) & \( 2 n_2 \partial^2_{xy} u_x+n_3 \partial^2_x u_y-2 \partial^2_y u_y-2 \partial^2_{yz} u_z \) \\ \hline
\( S_{5} \) & \( 2 n_1 \partial_x u_x+n_{-1} \partial_y u_y-4 \partial_z u_z \) \\ \hline
\( S_{6} \) & \( (2 \partial_y u_x+3 \partial_x u_y) \partial_x T+\partial_z T \partial_z u_y \) \\ \hline
\( S_{7} \) & \( 2 n_2 \partial^2_{xz} u_x-2 \partial^2_{yz} u_y+n_3 \partial^2_x u_z-2 \partial^2_z u_z \) \\ \hline
\( S_{8} \) & \( 2 n_1 \partial_x u_x-4 \partial_y u_y+n_{-1} \partial_z u_z \) \\ \hline
\( S_{9} \) & \( (2 \partial_z u_x+3 \partial_x u_z) \partial_x T+\partial_y T \partial_y u_z \) \\ \hline
\( S_{10} \) & \( -2 \partial^2_x u_x+n_3 \partial^2_y u_x+2 n_2 \partial^2_{xy} u_y \) \\ \hline
\( S_{11} \) & \( n_{-1} \partial_x u_x+2 n_1 \partial_y u_y-4 \partial_z u_z \) \\ \hline
\( S_{12} \) & \( (3 \partial_y u_x+2 \partial_x u_y) \partial_y T+\partial_z T \partial_z u_x \) \\ \hline
\( S_{13} \) & \( \partial^2_{yz} u_x+\partial^2_{xz} u_y+\partial^2_{xy} u_z \) \\ \hline
\( S_{14} \) & \( (\partial_z u_y+\partial_y u_z) \partial_x T+(\partial_x u_z+\partial_z u_x) \partial_y T+(\partial_x u_y+\partial_y u_x) \partial_z T \) \\ \hline
\( S_{15} \) & \( -2 \partial^2_x u_x+n_3 \partial^2_z u_x-2 \partial^2_{xy} u_y+2 n_2 \partial^2_{xz} u_z \) \\ \hline
\( S_{16} \) & \( n_{-1} \partial_x u_x-4 \partial_y u_y+2 n_1 \partial_z u_z \) \\ \hline
\( S_{17} \) & \( (3 \partial_z u_x+2 \partial_x u_z) \partial_z T+\partial_y T \partial_y u_x \) \\ \hline
\( S_{18} \) & \( -2 \partial^2_{xy} u_x+n_1 \partial^2_y u_y-2 \partial^2_{yz} u_z \) \\ \hline
\( S_{19} \) & \( -4 \partial_x u_x+(5+3 n) \partial_y u_y-4 \partial_z u_z \) \\ \hline
\( S_{20} \) & \( \partial_x T \partial_x u_y+\partial_z T \partial_z u_y \) \\ \hline
\( S_{21} \) & \( -2 \partial^2_{xz} u_x+2 n_2 \partial^2_{yz} u_y+n_3 \partial^2_y u_z-2 \partial^2_z u_z \) \\ \hline
\( S_{22} \) & \( -4 \partial_x u_x+2 n_1 \partial_y u_y+n_{-1} \partial_z u_z \) \\ \hline
\( S_{23} \) & \( (3 \partial_y u_z+2 \partial_z u_y) \partial_y T+\partial_x T \partial_x u_z \) \\ \hline
\( S_{24} \) & \( -2 \partial^2_{xy} u_x-2 \partial^2_y u_y+n_3 \partial^2_z u_y+2 n_2 \partial^2_{yz} u_z \) \\ \hline
\( S_{25} \) & \( -4 \partial_x u_x+n_{-1} \partial_y u_y+2 n_1 \partial_z u_z \) \\ \hline
\( S_{26} \) & \( (3 \partial_z u_y+2 \partial_y u_z) \partial_z T+\partial_x T \partial_x u_y \) \\ \hline
\( S_{27} \) & \( -2 \partial^2_{xz} u_x-2 \partial^2_{yz} u_y+n_1 \partial^2_z u_z \) \\ \hline
\( S_{28} \) & \( -4 \partial_x u_x-4 \partial_y u_y+(5+3 n) \partial_z u_z \) \\ \hline
\( S_{29} \) & \( \partial_x T \partial_x u_z+\partial_y T \partial_y u_z \) \\ \hline
\( S_{30} \) & \( n_2 \partial^2_x \rho-\partial^2_y \rho-\partial^2_z \rho \) \\ \hline
\( S_{31} \) & \( n_8 \partial^2_x T-\partial^2_y T-\partial^2_z T \) \\ \hline
\( S_{32} \) & \( -n_2(n^2+14 n+9)(\partial_x u_x)^2+(n^2+6 n-15)[(\partial_y u_y)^2+(\partial_z u_z)^2] \) \\ \hline
\( S_{33} \) & \( 4(n^2+12 n+15)(\partial_y u_y+\partial_z u_z) \partial_x u_x-8 n_9 \partial_y u_y \partial_z u_z \) \\ \hline
\( S_{34} \) & \( (n^2+11 n+20)[(\partial_y u_x)^2+(\partial_z u_x)^2]+n_{-1}[(\partial_x u_y)^2+(\partial_x u_z)^2]-n_7[(\partial_z u_y)^2+(\partial_y u_z)^2] \) \\ \hline
\( S_{35} \) & \( 4 n_3(\partial_y u_x \partial_x u_y+\partial_z u_x \partial_x u_z) \) \\ \hline
\( S_{36} \) & \( (4 n+11)(\partial_x T)^2+n_2(\partial_y T)^2+n_2(\partial_z T)^2 \) \\ \hline
\( S_{37} \) & \( n_2(\partial_x \rho)^2-(\partial_y \rho)^2-(\partial_z \rho)^2 \) \\ \hline
\( S_{38} \) & \( \rho \partial^2_{xy} \rho-\partial_x \rho \partial_y \rho \) \\ \hline
\( S_{39} \) & \( T \partial^2_{xy} T+3 \partial_x T \partial_y T \) \\ \hline
\( S_{40} \) & \( 2 n_{-5} \partial_x u_x+n_{11}(n_1 \partial_y u_y-2 \partial_z u_z) \) \\ \hline
\( S_{41} \) & \( (n^2+12 n+11) \partial_x u_x+2 n_{-5} \partial_y u_y-2 n_{11} \partial_z u_z \) \\ \hline
\( S_{42} \) & \( (n_9 \partial_z u_y+2 \partial_y u_z) \partial_z u_x+2 \partial_z u_y \partial_x u_z+2 \partial_y u_z \partial_x u_z \) \\ \hline
\( S_{43} \) & \( \rho \partial^2_{xz} \rho-\partial_x \rho \partial_z \rho \) \\ \hline
\( S_{44} \) & \( T \partial^2_{xz} T+3 \partial_x T \partial_z T \) \\ \hline
\( S_{45} \) & \( 2 n_{-5} \partial_x u_x+n_{11}(-2 \partial_y u_y+n_1 \partial_z u_z) \) \\ \hline
\( S_{46} \) & \( (n^2+12 n+11) \partial_x u_x-2 n_{11} \partial_y u_y+2 n_{-5} \partial_z u_z \) \\ \hline
\( S_{47} \) & \( (n_9 \partial_y u_z+2 \partial_z u_y) \partial_y u_x+2 \partial_x u_y \partial_y u_z+2 \partial_z u_y \partial_x u_y \) \\ \hline
\( S_{48} \) & \( -\partial^2_x \rho+n_2 \partial^2_y \rho-\partial^2_z \rho \) \\ \hline
\( S_{49} \) & \( -\partial^2_x T+n_8 \partial^2_y T-\partial^2_z T \) \\ \hline
\( S_{50} \) & \( -n_2(n^2+14 n+9)(\partial_y u_y)^2+(n^2+6 n-15)[(\partial_x u_x)^2+(\partial_z u_z)^2] \) \\ \hline
\( S_{51} \) & \( 4(n^2+12 n+15)(\partial_x u_x+\partial_z u_z) \partial_y u_y-8 n_9 \partial_x u_x \partial_z u_z \) \\ \hline
\( S_{52} \) & \( (n^2+11 n+20)[(\partial_x u_y)^2+(\partial_z u_y)^2]+n_{-1}[(\partial_y u_x)^2+(\partial_y u_z)^2]-n_7[(\partial_z u_x)^2+(\partial_x u_z)^2] \) \\ \hline
\( S_{53} \) & \( 4 n_3(\partial_y u_x \partial_x u_y+\partial_z u_y \partial_y u_z) \) \\ \hline
\( S_{54} \) & \( n_2(\partial_x T)^2+(4 n+11)(\partial_y T)^2+n_2(\partial_z T)^2 \) \\ \hline
\( S_{55} \) & \( -(\partial_x \rho)^2+n_2(\partial_y \rho)^2-(\partial_z \rho)^2 \) \\ \hline
\( S_{56} \) & \( \rho \partial^2_{yz} \rho-\partial_y \rho \partial_z \rho \) \\ \hline
\( S_{57} \) & \( T \partial^2_{yz} T+3 \partial_y T \partial_z T \) \\ \hline
\( S_{58} \) & \( 2 n_{-5} \partial_y u_y+n_{11}(-2 \partial_x u_x+n_1 \partial_z u_z) \) \\ \hline
\( S_{59} \) & \( -2 n_{11} \partial_x u_x+(n^2+12 n+11) \partial_y u_y+2 n_{-5} \partial_z u_z \) \\ \hline
\( S_{60} \) & \( (n_9 \partial_x u_z+2 \partial_z u_x) \partial_x u_y+2 \partial_y u_x \partial_x u_z+2 \partial_y u_x \partial_z u_x \) \\ \hline
\( S_{61} \) & \( -\partial^2_x \rho-\partial^2_y \rho+n_2 \partial^2_z \rho \) \\ \hline
\( S_{62} \) & \( -\partial^2_x T-\partial^2_y T+n_8 \partial^2_z T \) \\ \hline
\( S_{63} \) & \((n^2+6 n-15)[(\partial_x u_x)^2+(\partial_y u_y)^2]-n_2(n^2+14 n+9)(\partial_z u_z)^2 \) \\ \hline
\( S_{64} \) & \( 4(n^2+12 n+15)(\partial_x u_x+\partial_y u_y) \partial_z u_z-8 n_9 \partial_x u_x \partial_y u_y \) \\ \hline
\( S_{65} \) & \( (n^2+11 n+20)[(\partial_x u_z)^2+(\partial_y u_z)^2]+n_{-1}[(\partial_z u_x)^2+(\partial_z u_y)^2]-n_7[(\partial_y u_x)^2+(\partial_x u_y)^2] \) \\ \hline
\( S_{66} \) & \( 4 n_3(\partial_z u_x \partial_x u_z+\partial_z u_y \partial_y u_z) \) \\ \hline
\( S_{67} \) & \( n_2(\partial_x T)^2+n_2(\partial_y T)^2+(4 n+11)(\partial_z T)^2 \) \\ \hline
\( S_{68} \) & \( -(\partial_x \rho)^2-(\partial_y \rho)^2+n_2(\partial_z \rho)^2 \) \\ \hline
\end{longtable}

The expressions for the Third-order viscous stress $\bm{\Delta}_{2}^{*(3)}$
and heat flux $\bm{\Delta}_{3,1}^{*(3)}$ in three-dimensions are presented
in Table \ref{TableVI}, with the associated terms $S_{69}, \dots, S_{82}$
listed in Table \ref{TableVII}.

\begin{longtable}{|c|>{\centering\arraybackslash}m{12cm}|}
\caption{Expressions for $\bm{\Delta}_{2}^{*(3)}$ and $\bm{\Delta}_{3,1}^{*(3)}$.} \label{TableVI}\\ \hline
\textbf{TNE} & \textbf{Formula} \\ \hline
\endfirsthead

\hline
\textbf{TNE} & \textbf{Formula} \\ \hline
\endhead

\endlastfoot
\( \Delta_{2xx}^{*(3)} \) &
\( -2 n_2 n_3^{-2} \tau^3 \rho^{-2} R(n_3 \rho^2 R T^2 S_{69}+n_3 \rho^2 T \partial^2_x u_x S_{70}-n_3 \rho R T \partial^2_x \rho S_{71}+n_3 \rho^3 R \partial^2_x T S_{72}+\rho^2(\partial_x u_x)^2 S_{73}+n_3 \rho R \partial_x u_x S_{74}-n_3^2 R u_x \partial_x \rho S_{75}) \) \\ \hline
\( \Delta_{3,1x}^{*(3)} \) &
\( 2 n_3^{-2} \tau^3 \rho^{-2} R(n_3 \rho^2 R T^2 S_{76}-\frac{1}{4} n_3 \rho^2 R T \partial^2_x T S_{77}-\frac{1}{2} \rho^2 R T \partial^2_x u_x S_{78}-n_3 \rho R T^2 \partial^2_x \rho S_{79}-n_3 \rho^2 R(\partial_x T)^2 S_{80}-\frac{1}{4} \rho R T \partial_x T S_{81}+n_2 T S_{82}) \) \\ \hline
\end{longtable}

\begin{longtable}{|c|>{\centering\arraybackslash}m{12cm}|}
\caption{Expressions for $S_i$, $i=69,...,82$.} \label{TableVII}\\ \hline
\textbf{TNE} & \textbf{Formula} \\ \hline
\endfirsthead

\hline
\textbf{TNE} & \textbf{Formula} \\ \hline
\endhead

\endlastfoot
\( S_{69} \) & \( n_1 \rho \partial^3_x u_x-n_3 u_x \partial^3_x \rho \) \\ \hline
\( S_{70} \) & \( 2 n_{-1} \rho u_x \partial_x u_x+n_1 R T \partial_x \rho+(10 n+18) \rho R \partial_x T \) \\ \hline
\( S_{71} \) & \( (5 n+11) \rho T \partial_x u_x-2 n_3 u_x(T \partial_x \rho-\rho \partial_x T) \) \\ \hline
\( S_{72} \) & \( (5 n+7) T \partial_x u_x-2 n_3 u_x \partial_x T \) \\ \hline
\( S_{73} \) & \( 3(n^2-5) \rho T \partial_x u_x+n_{-1} n_3 u_x(T \partial_x \rho-\rho \partial_x T) \) \\ \hline
\( S_{74} \) & \( 2(5 n+9) \rho^2(\partial_x T)^2+(5 n+7) \rho T \partial_x \rho \partial_x T+(5 n+11) T^2(\partial_x \rho)^2 \) \\ \hline
\( S_{75} \) & \( (\partial_x \rho)^2 T^2-2 \rho T \partial_x \rho \partial_x T-\rho^2(\partial_x T)^2 \) \\ \hline
\( S_{76} \) & \( -\frac{1}{4}(n^2+12 n+23) \rho R \partial^3_x T-\frac{1}{2} n_{-1} \rho u_x \partial^3_x u_x+n_2 R T \partial^3_x \rho \) \\ \hline
\( S_{77} \) & \( 12(n^2+11 n+22) \rho R \partial_x T+(n^2+12 n+23) R T \partial_x \rho+2(n^2+10 n+13) \rho u_x \partial_x u_x \) \\ \hline
\( S_{78} \) & \( n_1 n_3 n_{11} \rho u_x \partial_x T+(20 n^2+52 n+12) \rho T \partial_x u_x+n_3 n_{-1} T u_x \partial_x \rho \) \\ \hline
\( S_{79} \) & \( -\frac{3}{4}(n^2+12 n+23) \rho R \partial_x T+n_2(2 R T \partial_x \rho-\rho u_x \partial_x u_x) \) \\ \hline
\( S_{80} \) & \( \frac{1}{4}(7 n^2+68 n+129) \rho R \partial_x T+(n^2+11 n+21) R T \partial_x \rho+\frac{1}{2}(n^2+12 n+17) \rho u_x \partial_x u_x \) \\ \hline
\( S_{81} \) & \( 2(3 n^3+62 n^2+197 n+158) \rho^2(\partial_x u_x)^2+2 n_3(n^2+10 n+13) \rho u_x \partial_x \rho \partial_x u_x+3 n_3(n^2+12 n+23) R T(\partial_x \rho)^2 \) \\ \hline
\( S_{82} \) & \( -3 n_1 \rho^2 R T \partial_x \rho(\partial_x u_x)^2+n_3 R^2 T^2(\partial_x \rho)^3-n_{-1} \rho^3 u_x(\partial_x u_x)^3-n_3 \rho R T u_x(\partial_x \rho)^2 \partial_x u_x \) \\ \hline
\end{longtable}

\section*{Declaration of interests}

The authors report no conflict of interest.

\section*{Author ORCIDs}

H. Lai, https://orcid.org/0000-0001-5978-5736

Q. Guo, https://orcid.org/0009-0004-9568-4712

Y. Gan, https://orcid.org/0000-0002-0191-9022

B. Yang, https://orcid.org/0000-0003-4015-199X

H. Liu, https://orcid.org/0000-0002-8780-0398

P. Lin, https://orcid.org/0000-0003-2361-0066


\bibliographystyle{jfm}
\bibliography{References}

\end{document}